\documentclass[11pt,a4paper]{article}
\usepackage{jinstpub}

\usepackage[english]{babel}

\usepackage{amsmath}
\usepackage{booktabs}
\usepackage{caption}
\usepackage{duckuments}
\usepackage{graphicx}
\usepackage{hyperref}
\usepackage{lineno}
\usepackage{multirow}
\usepackage{pdflscape}
\usepackage{placeins}
\usepackage{subcaption}
\usepackage[notlof,notlot,nottoc]{tocbibind}
\usepackage{todonotes}
\usepackage{units}
\usepackage{upgreek}
\usepackage{xspace}

\newcommand{\wrap}[1]{\ensuremath{#1}\xspace}
\newcommand{\ifb}{\wrap{\textrm{fb}^{-1}}}
\newcommand{\um}{\wrap{\muup\textrm{m}}}
\newcommand{\uA}{\wrap{\muup\textrm{A}}}
\newcommand{\oC}{\wrap{^{\circ}\textrm{C}}}
\newcommand{\COTwo}{\wrap{\textrm{CO}_2}}

\graphicspath{{figures/}}

\title{Cracking under pressure --- investigating mitigation approaches for silicon fractures on ATLAS strip tracker petals at cold temperatures}

\author[a]{S.H.~Abidi,}
\author[b]{J.-H.~Arling,}
\author[c,d,1]{M.J.~Basso,}\emailAdd{mbasso@triumf.ca}
\author[c,d]{S.~Beaupre,}
\author[b]{I.~Bloch,}
\author[e]{A.J.~Blue,}
\author[b]{M.~Caspar,}
\author[c,d]{J.~Chen,}
\author[b]{S.~D{\'i}ez~Cornell,}
\author[c,f]{U.~Epstein,}
\author[c,1]{E.K.~Filmer,}\emailAdd{efilmer@triumf.ca}
\author[c,d]{A.~Fournier,}
\author[b]{L.~Franconi,}
\author[f]{A.~Gabrielli,}
\author[b]{J.A.~Hallford,}
\author[b]{S.~Heim,}
\author[g]{C.M.~Helling,}
\author[c]{N.P.~Hessey,}
\author[b]{R.M.~Jacobs,}
\author[b]{T.~Kuhl,}
\author[c,f]{M.~Licht,}
\author[c,f,g,1]{C.K.~Mahajan,}\emailAdd{cmahajan@student.ubc.ca}
\author[c,d]{S.~Manson,}
\author[b]{K.~Mauer,}
\author[b]{S.~Oerdek,}
\author[h]{V.~Platero~Montagut,}
\author[c,d]{L.~Poley,}
\author[b]{C.O.~Sander,}
\author[b,2]{K.~Ran,}
\author[b]{S.~Sinha,}
\author[h]{C.~Solaz~Contell,}
\author[h]{U.~Soldevila~Serrano,}
\author[i]{D.~Sperlich,}
\author[c,d]{B.~Stelzer,}
\author[c]{A.H.~Tigchelaar,}
\author[h]{E.~Torres~Reoyo,}
\author[j]{G.~Vallone,}
\author[f]{and L.M.~Veloce}

\affiliation[a]{Physics Department, Brookhaven National Laboratory, Upton, NY, United States of America}
\affiliation[b]{Deutsches Elektronen-Synchrotron DESY, Hamburg and Zeuthen, Germany}
\affiliation[c]{TRIUMF, Vancouver, BC, Canada}
\affiliation[d]{Department of Physics, Simon Fraser University, Burnaby, BC, Canada}
\affiliation[e]{SUPA - School of Physics and Astronomy, University of Glasgow, Glasgow, United Kingdom}
\affiliation[f]{Department of Physics, University of Toronto, Toronto, ON, Canada}
\affiliation[g]{Department of Physics, University of British Columbia, Vancouver, BC, Canada}
\affiliation[h]{Instituto de F{\'i}sica Corpuscular (IFIC), Centro Mixto Universidad de Valencia - CSIC, Valencia, Spain}
\affiliation[i]{Physikalisches Institut, Albert-Ludwigs-Universit{\"a}t Freiburg, Freiburg, Germany}
\affiliation[j]{Physics Division, Lawrence Berkeley National Laboratory, Berkeley, CA, United States of America}

\note{Corresponding authors.}
\note{Now at Fudan University, Shanghai, China.}

\abstract{%

For the High-Luminosity upgrade of the Large Hadron Collider, the ATLAS experiment will replace its current Inner Detector with an all-silicon Inner Tracker (ITk), consisting of pixel and strip detectors. The strip detector will consist of a central region or ``barrel'' assembled with staves and forward regions or ``end-caps'' assembled with petals. The ITk will nominally operate with liquid \COTwo cooling at $\unit[-35]{\oC}$; however, in the event of cooling system failures, it is possible that sensors will experience temperatures below $\unit[-35]{\oC}$. At these low temperatures, it has been observed that the silicon sensors within modules --- the fundamental readout units of the detector --- can physically crack, rendering the modules inoperable. Understanding and resolving the issue of sensor cracking was one of the most important and urgent issues for the ITk project. This paper presents part of the mitigation strategies developed for petals. These mitigation strategies are based on modifications to the choice of adhesive and its deposition pattern for module assembly and petal loading. The most promising mitigation strategy presented here prevents cracking to temperatures as low as $\unit[-45]{\oC}$, which can be expected in case of cooling system problems, with a small percentage of cracks observed after being cycled to $\unit[-55]{\oC}$, which can be expected in case of catastrophic cooling system failures.

}


\begin{document}

\maketitle
\flushbottom

\section{Introduction}
\label{sec:intro}

After the completion of Run~3 of the Large Hadron Collider (LHC)~\cite{Evans:2008zzb}, the High-Luminosity LHC (HL-LHC)~\cite{ZurbanoFernandez:2020cco} is scheduled to begin operation in the middle of 2030. It will increase the size of the LHC dataset by an order of magnitude over its lifetime, up to \unit[3000]{\ifb} of integrated luminosity. The HL-LHC will deliver a peak instantaneous luminosity of $\unit[7.5\times 10^{34}]{cm^{-2}s^{-1}}$, which corresponds to approximately 200~proton-proton collisions per bunch crossing, up from an average of 60~collisions per bunch crossing in Run~3. Owing to these conditions, the track density and radiation dose will increase by an order of magnitude. To address the challenges of the dense collision environment of the HL-LHC, the ATLAS experiment~\cite{ATLAS:2008xda} will replace its current Inner Detector~\cite{ATLAS:1997ag, ATLAS:1997af} with an all-silicon Inner Tracker (ITk)~\cite{ATLASCollaboration:2012ilu}, featuring radiation-hard readout electronics, increased granularity, forward coverage up to a pseudorapidity $|\eta| = 4.0$, and minimal inactive material.

\begin{figure}[htbp]
    \centering
    \begin{subfigure}[b]{0.45\textwidth}
        \includegraphics[width=\textwidth]{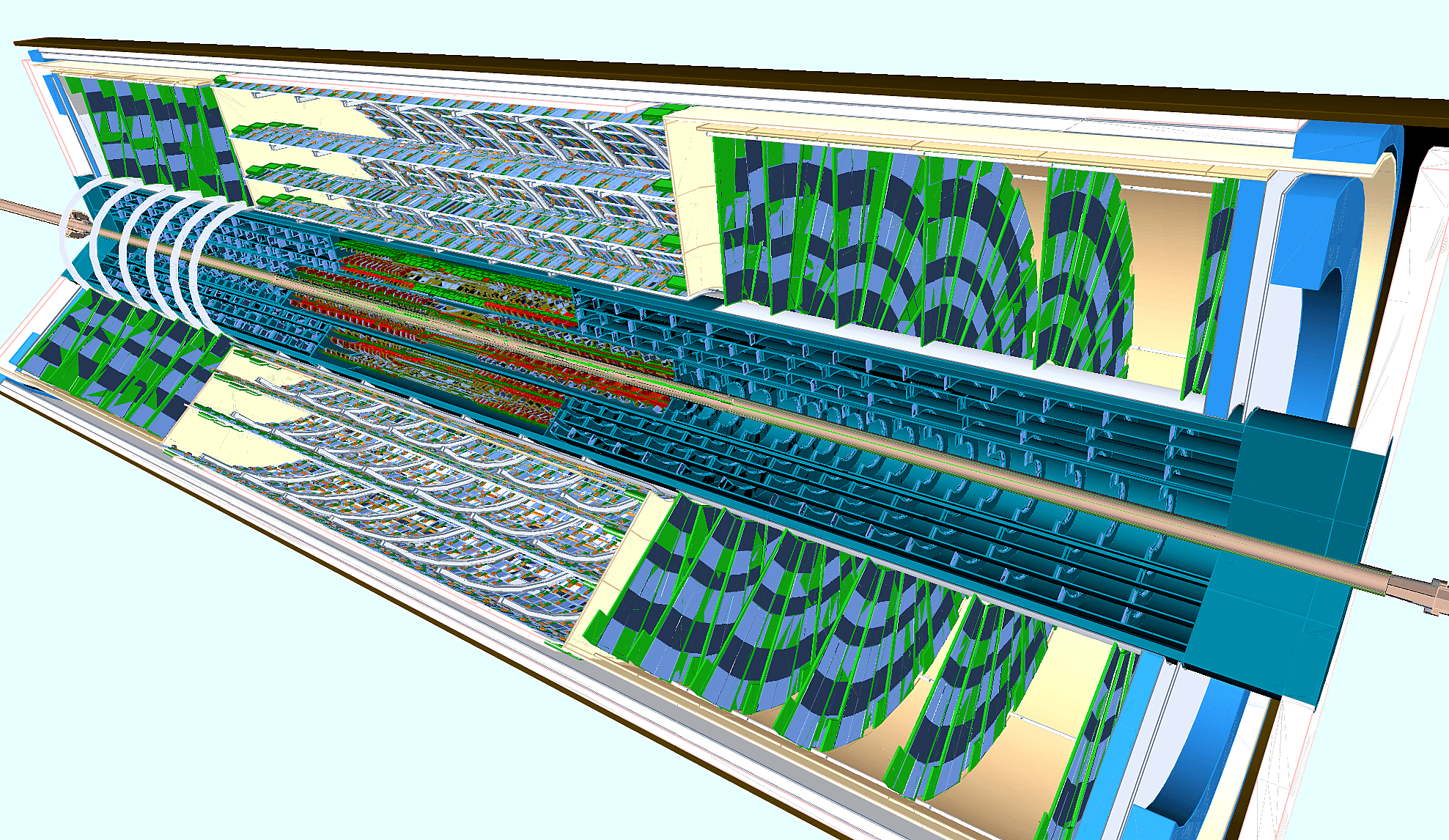}
        \vspace{3mm}
        \caption{}
        \label{fig:itk_overview}
    \end{subfigure}
    \hfill
    \begin{subfigure}[b]{0.54\textwidth}
        \includegraphics[width=\textwidth]{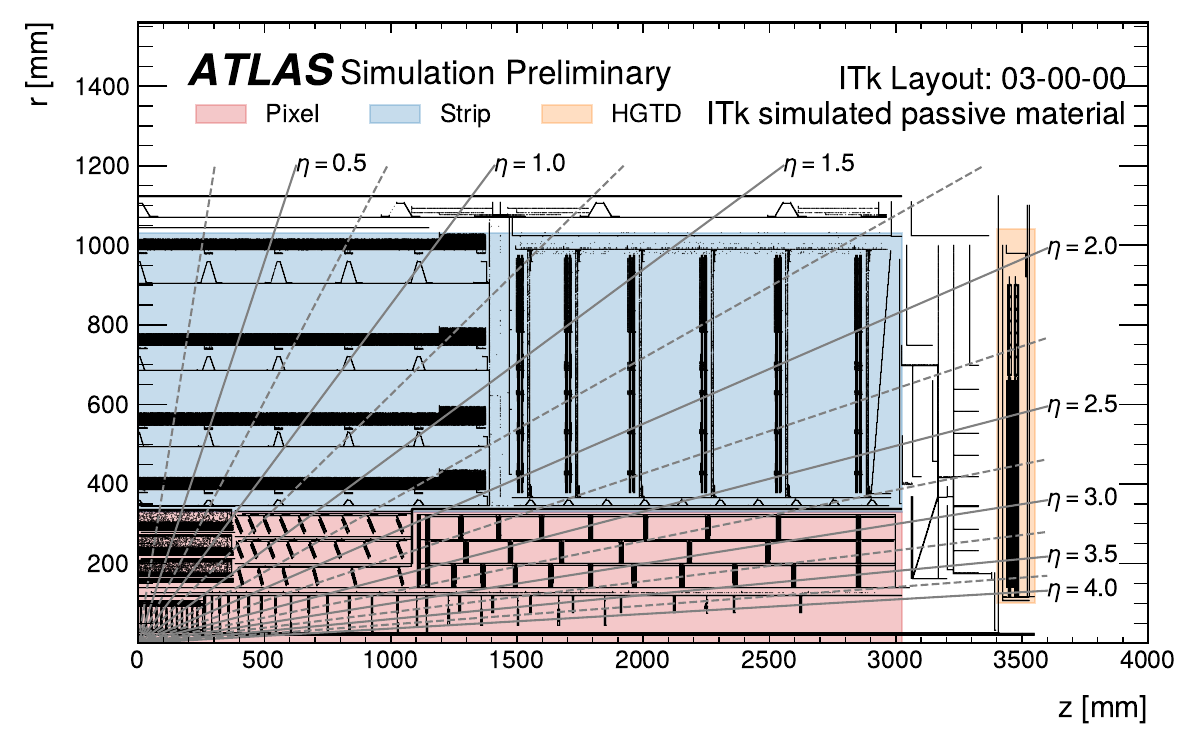}
        \caption{}
        \label{fig:itk_layout}
    \end{subfigure}
    \caption{(a) Visualization of the ITk including the inner pixel detector and outer strip detector~\cite{ATLAS:2017azf}. (b) Schematic overview of one quadrant of the ITk with the pixel system in red and the strip system in blue~\cite{ITK-2023-001}. The horizontal axis is parallel to the beam line with the origin at the interaction point, and the vertical axis is the perpendicular radius from the interaction point.}
    \label{fig:itk_main}
\end{figure}

\begin{figure}[htbp]
    \centering
    \begin{subfigure}{0.36\textwidth}
        \includegraphics[width=\textwidth]{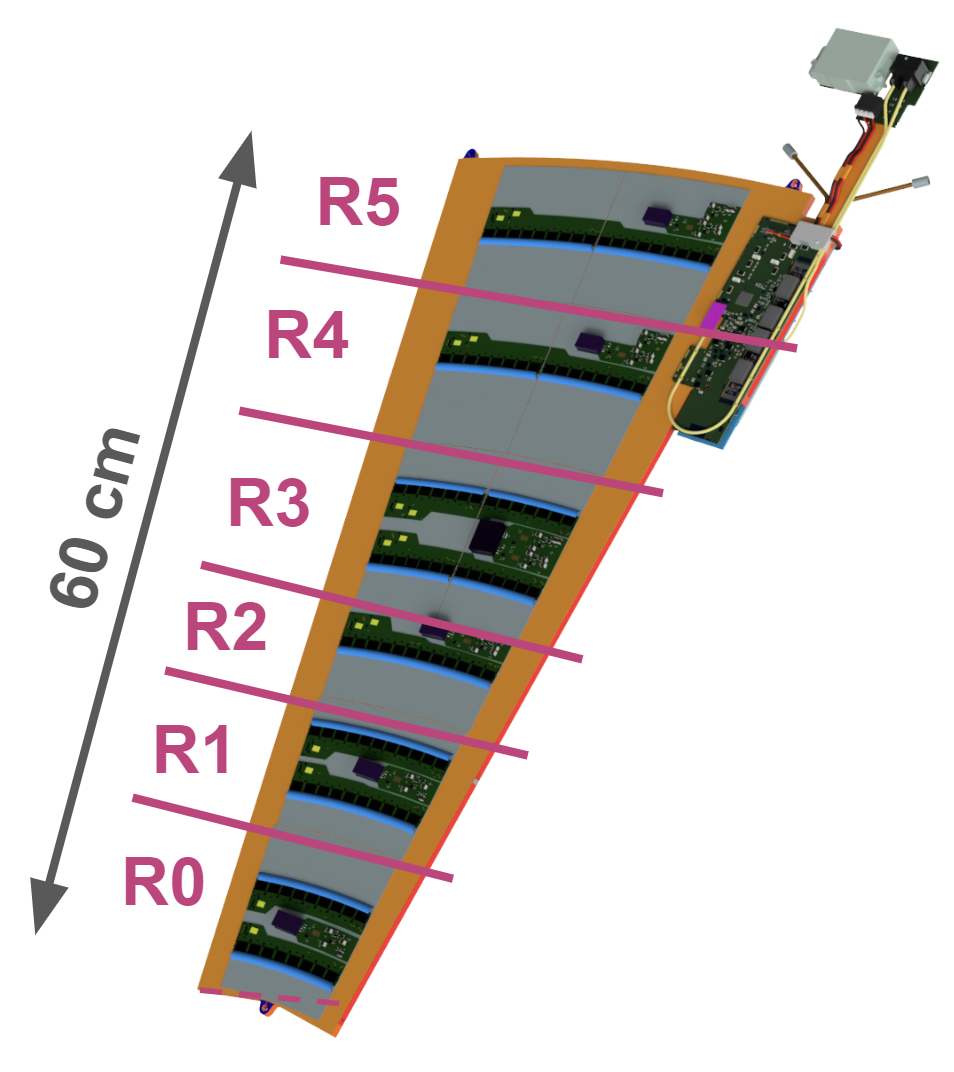}
        \caption{}
    \end{subfigure}
    \begin{subfigure}{0.56\textwidth}
        \includegraphics[width=\textwidth]{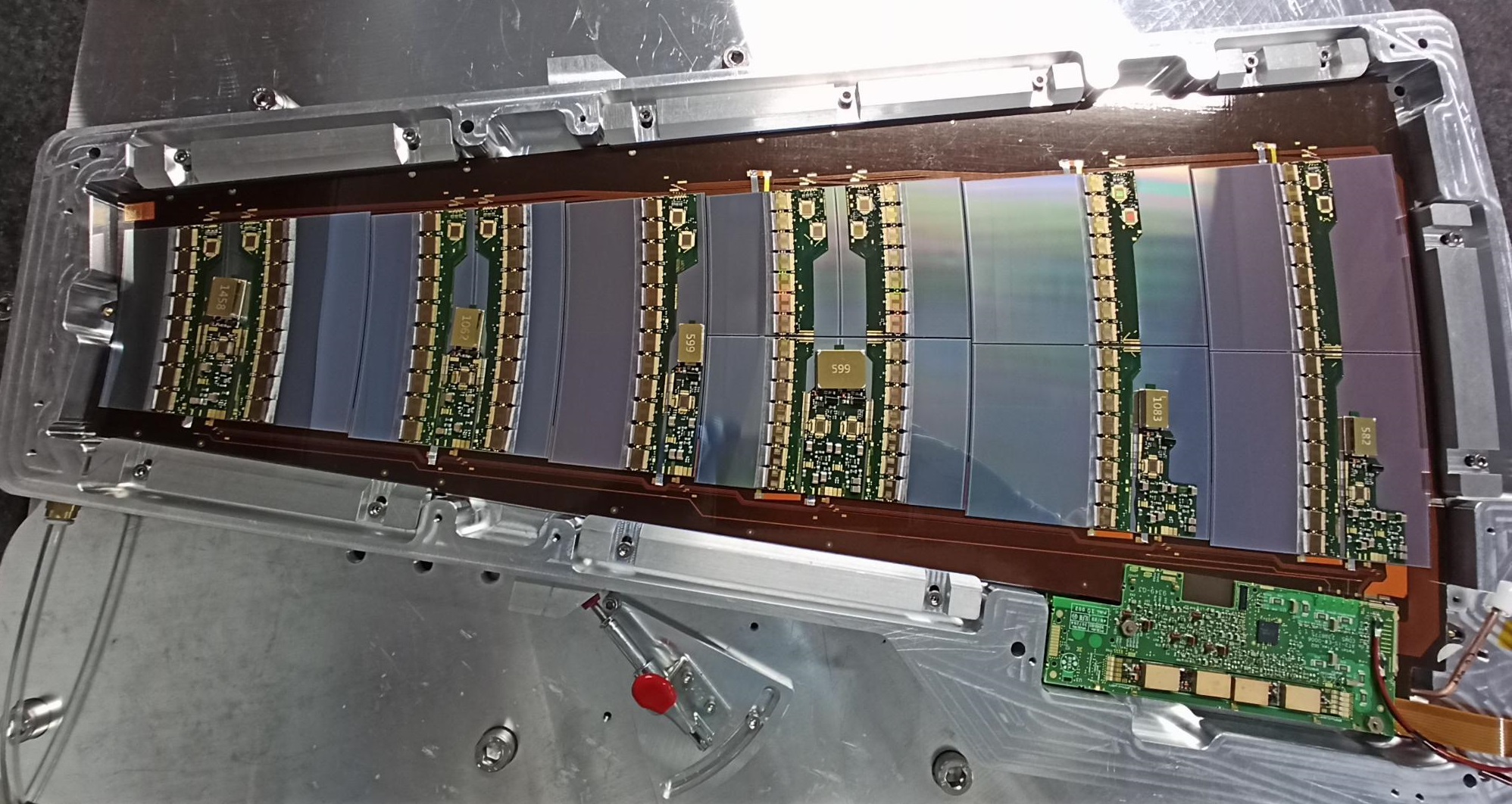}
        \caption{}
    \end{subfigure}
    \caption{(a) Schematic of petal with the different module types (R0--R5) delineated~\cite{Basso:2025hub}. (b) Picture of a loaded petal.}
    \label{fig:petal}
\end{figure}

The ITk is pictured in figure~\ref{fig:itk_overview}: it is composed of a pixel detector~\cite{ATLAS:2017svb}, located closest to the beam line, and a strip detector~\cite{ATLAS:2017azf} surrounding the pixel detector. The strip detector consists of a barrel region with four cylindrical layers and two end-cap regions each with six disks, as shown in figure~\ref{fig:itk_layout}. The strip detector is built of two local support structures: staves for the barrel region and petals for the end-cap regions~\cite{Diez:2024bvt}; this paper will focus on petals. A petal, seen in figure~\ref{fig:petal}, is made of a carbon-fibre core of a radial shape. Copper-on-polyimide bus tapes on the surface of the core route data and power. Titanium pipes embedded within a thermally-conductive foam comprise the interior of the core and provide cooling down to $\unit[-35]{\oC}$. The petal is double sided, with a main side and a secondary side. Twelve silicon modules~\cite{ATLAS:2020ize, Kuehn:2017cqo} --- the fundamental readout units of the strip detector --- are glued or ``loaded'' onto the core, six on each side, using an automated assembly system~\cite{Stelzer:2024hwq}. The automated assembly system is capable of dispensing adhesive and placing modules with micron-level precision. The End-of-Substructure (EoS) cards~\cite{Boebel:2024dcu} on a petal serve as the interface between the modules on that petal and the off-detector systems.

Owing to the radial shape of the petal, each of the six modules on a single side is of a unique type and labelled R0 through R5 in order of increasing radius. The module types offer the same functionality and differ only in their geometry. A schematic of a module is shown in figure~\ref{fig:module_schematic}. Each module consists of one or two silicon strip sensors~\cite{Unno:2023bot}, onto which are glued printed circuit boards (PCBs) known as hybrids and powerboards~\cite{Haber:2023kha}. The hybrids perform readout via front-end ATLAS Binary Chips (ABCs)~\cite{Basso:2022qob} and control via Hybrid Control Chips (HCCs)~\cite{ATLAS:2023fwl}. The powerboards provide low voltage (LV) --- accomplished using a DC-DC converter --- and high voltage (HV) to the readout electronic and sensors, respectively, and monitor conditions via Autonomous Monitoring and Control chips (AMACs)~\cite{Gosart:2023pcl}. The design of the hybrids and powerboards also vary for each module type to adapt to the geometry. Each module has between one and four hybrids of different types, labelled H0 through H3.\footnote{For example, ``R0H0'' corresponds to the H0 hybrid of an R0 module, while ``R3H2'' corresponds to the H2 hybrid of an R3 module.} On modules with two physical sensors or ``split'' modules --- R3, R4, and R5 modules --- the right and left halves are labelled by M0 and M1, respectively.\footnote{For example, ``R4M0'' corresponds to the right half of an R4 module.} Modules undergo extensive quality control (QC)~\cite{Tishelman-Charny:2024yys}. Of particular importance is the ability to reach an HV bias voltage of $\unit[-500]{V}$ with sensor leakage current and noise performance within acceptance criteria (described in section~\ref{subsec:PetalQC}).

\subsection{Sensor cracking}
\label{sec:sensor_cracking}

\begin{figure}[htbp]
    \centering
    \begin{subfigure}[b]{0.55\textwidth}
         \includegraphics[width=\textwidth]{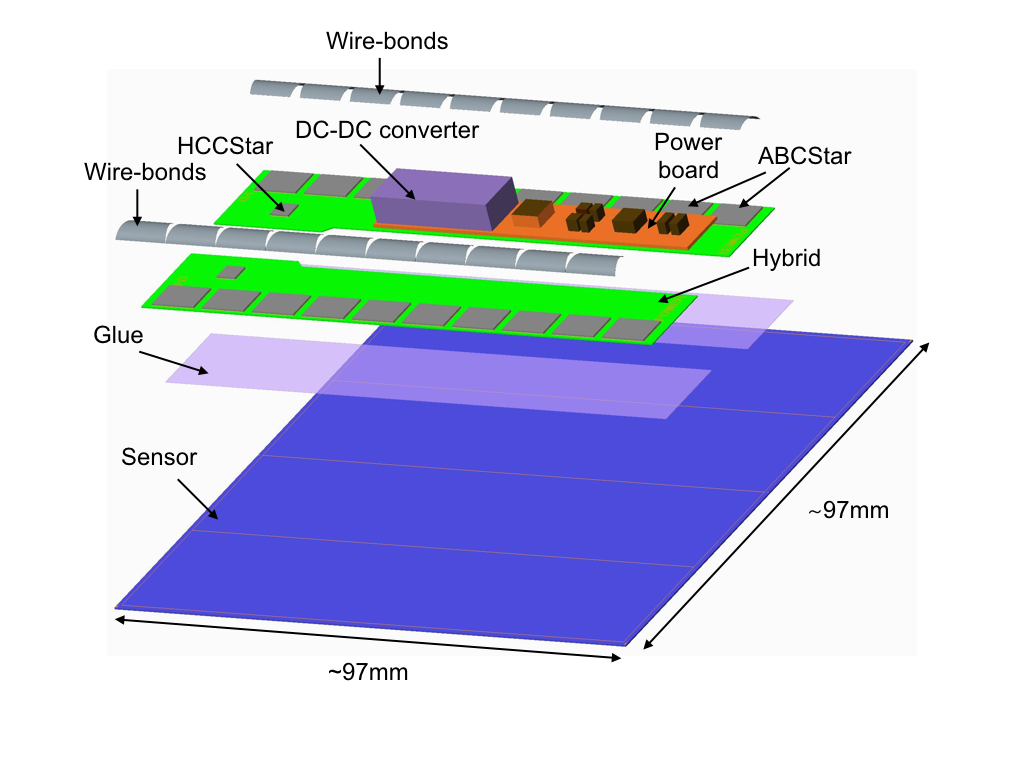}
         \caption{}
         \label{fig:module_schematic}
    \end{subfigure}
    \begin{subfigure}[b]{0.44\textwidth}
        \includegraphics[width=\textwidth]{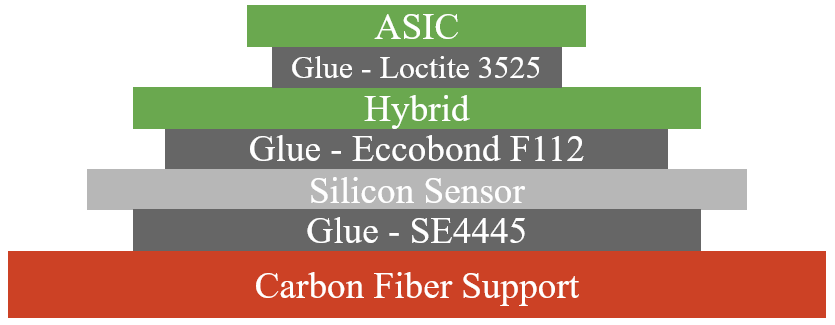}
        \caption{}
        \label{fig:module_glue_schematic}
    \end{subfigure}
    \caption{(a) Schematic of a silicon strip sensor module~\cite{ATLAS:2017azf}. While the module shown is for use in the barrel detector, the end-cap modules share the same basic structure and components. (b) Schematic of layers within a module loaded onto a petal. An adhesive with a low Young's modulus, SE4445, is used to load the module to the petal; within the module, an adhesive with a high Young's modulus, Eccobond F112, is used to attach hybrids and powerboards to the silicon sensor.}
    \label{fig:module}
\end{figure}

During the pre-production of petals and staves, it was found that several modules loaded onto cores failed electrical tests~\cite{Tishelman-Charny:2024wxu, Basso:2025hub} after being tested at $\unit[-25]{\oC}$ (staves) or colder (petals) as part of petal/stave QC, despite passing module QC requirements prior to loading. These modules exhibited runaway leakage currents at bias voltages below their operational requirements. They were also found to have localized regions of low or high noise. Upon visual inspection of some of these regions, physical cracks of the silicon sensors were observed, predominantly in the sensor areas under and between PCBs. Simulations of the detector~\cite{simulation} suggest that cracks are caused by a mismatch in the coefficients of thermal expansion (CTEs) between the copper-based PCBs and the silicon sensor. A cross section of a loaded module is shown in figure~\ref{fig:module_glue_schematic}. A silicone gel, SE 4445 CV (``SE4445'')~\cite{se4445:tds}, is used below the sensor and a stiff adhesive, Eccobond F112~\cite{trueblue:tds}, is used to attach the PCBs directly onto the silicon sensor. This combination leads to areas of mechanical stress within the sensor at cold temperatures, where the copper inside the PCBs contracts more than the silicon, resulting in cracks. Dedicated measurements have shown that the majority of sensors only crack for stresses above \unit[350]{MPa}, but outliers can show cracks at \unit[200]{MPa} and extreme outliers can show cracks as early as \unit[100]{MPa}~\cite{Abidi:2025znz}.

Sensor cracking was one of the most urgent issues within the ITk as it renders a full module unusable for tracking. While exposure to temperatures below $\unit[-35]{\oC}$ should not occur during normal operation, temperatures of $\unit[-45]{\oC}$ or in extreme cases $\unit[-55]{\oC}$ are possible during a power or cooling system failure, the latter only occurring in case of actual physical damage of the ITk cooling system. These scenarios are more likely to occur during the surface integration phase of the experiment. To address sensor cracking, two mitigation strategies have been explored by the ATLAS ITk community. The first strategy focuses on modifications to the choice and pattern of the adhesive used during module assembly and petal loading, requiring minimal changes to the assembly process (``Improved Nominal''). The second strategy involves the inclusion of flexible kapton layer (or ``interposer'') between the sensor and PCBs; this strategy is discussed in detail elsewhere~\cite{Fortman:2025ocz, DAmen:2025grc} and is therefore not subject of this publication.

This paper focuses on the first strategy and studies the performance for petals assembled and tested at several assembly institutes. Its structure is as follows: section~\ref{sec:setup} describes the experimental setup at the various sites and outlines the electrical tests conducted to detect cracked silicon; section~\ref{sec:results} presents the results for eight petals assembled at various stages of the investigation into sensor cracking; and section~\ref{sec:conclusion} summarizes the conclusions of the study.

\FloatBarrier
\section{Experimental setup and methodology}
\label{sec:setup}

\begin{figure}[htbp]
    \centering
    \includegraphics[width=0.55\textwidth]{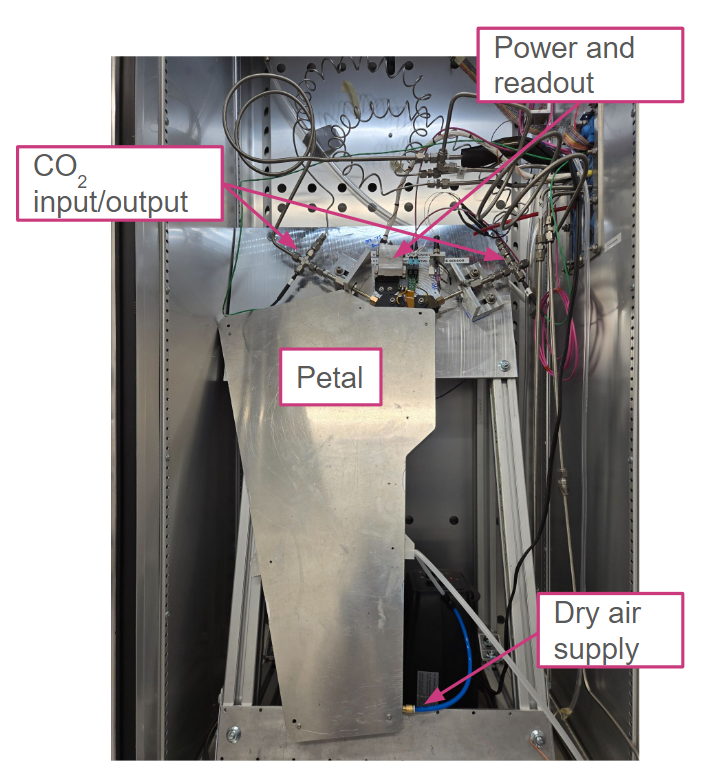}
    \caption{Labelled picture of a petal in the enclosure for electrical testing~\cite{Basso:2025hub}.}
    \label{fig:setup}
\end{figure}

To verify that a petal meets its design requirements, extensive electrical tests are conducted during the QC process. The testing setup at each of the four ITk petal loading sites consists of a light-tight, thermally-isolated enclosure to hold the petal. Both the enclosure and the petal case are flushed with dry air to keep the relative humidity low, $< 10\%$, to prevent early breakdown of sensors~\cite{Fadeyev:2024ubo}. Three of the four loading sites connect the petal to an active dual-phase \COTwo cooling plant, the MARTA system~\cite{MARTA}, which allows for testing at temperatures ranging from +20 to $\unit[-35]{\oC}$. The fourth institute uses an active mono-phase ethanol cooling plant connected to the petal which allows for testing at temperatures ranging from +20 to $\unit[-75]{\oC}$. Both cooling systems allow for testing at temperatures at which the ITk is expected to operate, and the former is reflective of what will be used during operation, +20 to $\unit[-35]{\oC}$~\cite{Diez:2024zpb}. Figure~\ref{fig:setup} shows a picture of a petal connected to a liquid \COTwo system, set up inside an enclosure.

The readout electronics for the petal and its modules are powered using $\unit[11]{V}$ from an LV supply, and an HV supply capable of providing $\unit[-550]{V}$ or lower is used to bias each sensor. Additionally, each module has a built-in HV multiplexer for toggling the voltage applied to its sensors to enable independent control of modules connected to the same HV channel~\cite{Villani:2020mqp}.

Communication with the petal and its modules is instrumented using $\unit[10]{Gbps}$ optical readout. Electrical-to-optical conversion occurs on the EoS~\cite{Troska:2017eur, Moreira:2025iiq}, which is interfaced using fibre optics with a Genesys~2 FPGA development board~\cite{DigilentG2}. The FPGA is read out over fibre optics or $\unit[1]{Gbps}$ ethernet using the Inner Tracker Strips Data Acquisition (ITSDAQ) software~\cite{ATLAS:2020ize} and firmware~\cite{ITSDAQfw}.

\subsection{Petal quality control}
\label{subsec:PetalQC}

Pre-production petal testing begins with non-electrical and electrical QC. Non-electrical QC is not a major consideration for this paper and is described elsewhere~\cite{Stelzer:2024hwq}. Electrical QC begins with an initial current-voltage (IV) scan for each module (section~\ref{subsubsec:iv}) followed by a set of data acquisition (DAQ) scans for each strip of each module, including measurements of the noise, gain, and offset (section~\ref{subsubsec:daq}). These initial tests are conducted at both warm ($\unit[+15]{\oC}$) and cold ($\unit[-35]{\oC}$) temperatures. After the initial set of tests, thermal cycling is performed, progressively reaching colder temperatures (section~\ref{subsubsec:tc}). Note that under normal petal QC, thermal cycling only reaches $\unit[-35]{\oC}$. However, the petals assembled for cracking mitigation tests were subjected to thermal cycles as low as $\unit[-75]{\oC}$. Between each temperature step, IV and DAQ scans are repeated at warm --- and sometimes cold --- temperatures.

\subsubsection{Current-voltage scans or IV test}
\label{subsubsec:iv}

An IV test is performed for each module before the first thermal cycle and again after each subsequent cycle to assess the performance of the sensor. The bias voltage is incremented from 0 to $\unit[-550]{V}$ in $\unit[-10]{V}$ steps, with a dwell time of \unit[10]{s} per step. At each step, the leakage current of each module is measured. Two independent current compliance limits are applied during the measurement. The first compliance limit is set by the AMAC and requires at most $\unit[10]{\uA}$ per sensor. The second, moving compliance limit is set by the HV power supply. An additional QC requirement specifies that the leakage current for each sensor must remain below $\unit[0.1]{\uA/cm^2}$ at $\unit[-500]{V}$ to ensure that the detector can be adequately powered~\cite{Klein:2024hen}. Given that the QC requirement is applied after the test finishes, it is possible for a module to complete an IV scan without reaching either compliance limit and still fail the test.

\begin{figure}[htbp]
    \centering
    \includegraphics[width=\textwidth]{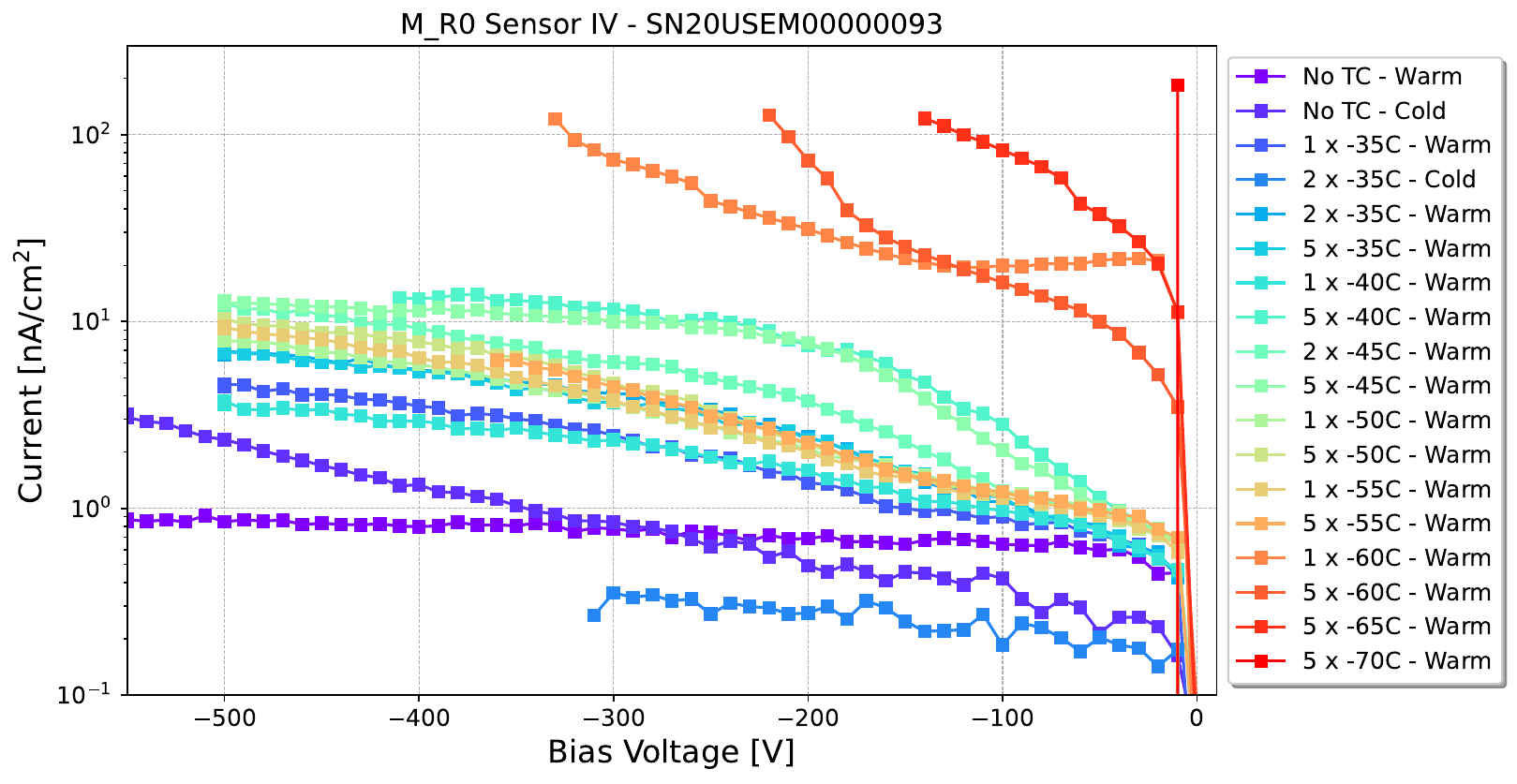}
    \caption{IV scans for an R0 module on a petal (``VAN3''); results are shown for various scans. In the legend, ``$N \times Y\textrm{C}$'' indicates a scan after $N$ thermal cycles at a temperature of $\unit[Y]{\oC}$. ``Warm'' and ``Cold'' indicate scans performed at +20 and $\unit[-35]{\oC}$, respectively.}
    \label{fig:worsing_iv_example}
\end{figure}

A rapid, exponential increase in sensor leakage current --- known as early breakdown --- could be indicative of sensor damage, damage from electrostatic discharge, contamination, or cracks. It is therefore considered one of the indicators of a sensor crack, but not conclusive proof. Figure~\ref{fig:worsing_iv_example} shows examples of IV scans. Scans ending at $\unit[-500]{V}$ or $\unit[-550]{V}$ correspond to normal IV scans. Scans ending before $\unit[-500]{V}$ and below $\unit[100]{nA/cm^2}$ were terminated due to the compliance limit set by the AMAC. Scans reaching $\unit[100]{nA/cm^2}$ were terminated due to the compliance limit set by the HV power supply.

\subsubsection{Data acquisition scans}
\label{subsubsec:daq}

Similar to IV scans, DAQ threshold scans are performed for each module before the first thermal cycle and again after each subsequent cycle to assess the performance of the readout electronics~\cite{ATLAS:2020ize}. For a given module, these scans repeatedly inject known charges using the built-in calibration circuit of the ABCs and read out the module to extract signal-above-threshold (``hit'') efficiencies as a function of some tunable parameter, such as the threshold of the discriminator in front-end circuitry or the latency.

Of particular importance is the measurement of the gain and output noise of the front-end circuitry for a given input charge, yielding a measurement of the input noise. The output noise is defined as the noise at the input of the discriminator, while the input noise is defined as the noise at the input of the preamplifier and is calculated as the output noise divided by the gain~\cite{Cormier:2021oog}. The QC requirement for the maximum input noise is set by the required signal-to-noise ratio of 10:1 at end-of-life for the ITk. The expected input noise is estimated based on the module geometry and sensor characteristics. The input noise also varies as a function of temperature and bias voltage, where lower noise is expected at a lower temperature and higher noise is expected for a smaller bias voltage.

Input noise is measured for each sensor strip --- each an independent readout channel --- of a given module. These channels are split into two ``streams'': those which extend under the PCBs, referred to as the ``under'' stream, and those which do not, referred to as the ``away'' stream. This is exemplified by figure~2 of ref.~\cite{Unno:2023bot}, where the different rows of sensor strips correspond to independent streams when bonded to an ABC. In general, the input noise measured for the channels in the under stream is higher than that measured for the channels in the away stream (for comparable strip lengths) due to capacitive coupling of the sensors to the adhesive and the PCBs.

\begin{figure}[htbp]
    \centering
    \includegraphics[width=\textwidth]{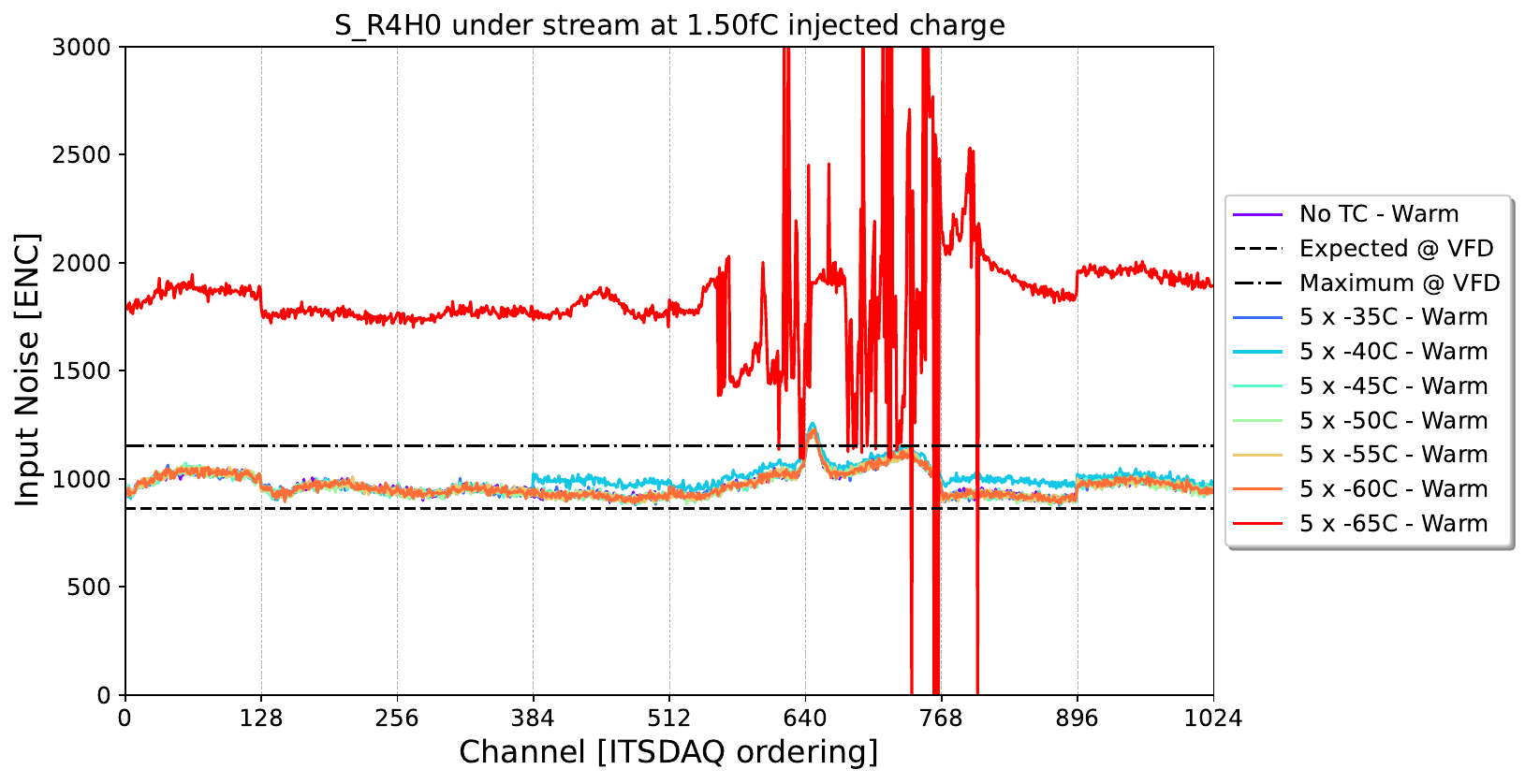}
    \caption{Typical noise signature from a crack on an R4 module on a petal (``VAN3''). Results are shows for every fifth thermal cycle and were each obtained at $\unit[+20]{\oC}$ (``Warm''). Also shown are the expected and maximum noise margins at full depletion voltage.}
    \label{fig:worsing_daq_example}
\end{figure}

Note that throughout this paper and unless stated otherwise, noise results are quoted for an input calibration charge of $\unit[1.5]{fC}$ at a bias voltage of $\unit[-350]{V}$, as this is where the sensors are, on average, fully depleted. Units of $\unit[1]{ENC} = \unit[1.6 \times 10^{-4}]{fC}$ are used for input noise. However, if early breakdown occurs for the corresponding sensor, noise results may be obtained for a bias voltage below $\unit[-350]{V}$. Figure~\ref{fig:worsing_daq_example} shows example input noise results from DAQ scans of an R4 module with a crack. The scans before thermal cycling to $\unit[-65]{\oC}$ are indicative of a properly functioning module. The scan after thermal cycling to $\unit[-65]{\oC}$ has a much higher average noise value due to an applied bias voltage less than the full depletion voltage.

\subsubsection{Thermal cycling}
\label{subsubsec:tc}

Sensor cracking was first observed on petals following testing at $\unit[-35]{\oC}$, and it is imperative that these objects are exposed to as cold or colder temperatures to understand the effectiveness of a particular mitigation strategy. Exposure to these cold temperatures is realized using thermal cycling. Starting at $\unit[+20]{\oC}$, the cooling system reduces the temperature of the petal to a cold set-point, $\leq \unit[-35]{\oC}$, before returning to $\unit[+20]{\oC}$ at a ramp rate of up to $\unit[\pm 2.5]{\oC}$ per minute, depending on the cooling system. This constitutes a single thermal cycle. Between one and five thermal cycles are repeated for each set-point prior to the testing. After completing five thermal cycles, the set-point temperature is lowered by $\unit[5]{\oC}$ and the procedure is repeated. Thermal cycling is performed for set-points as low as $\unit[-75]{\oC}$.

As the MARTA system only cools to $\unit[-35]{\oC}$, an external climate chamber is used to thermal cycle the petal. During climate chamber cycles, the petal is not powered and no electrical tests can be done. The alternative ethanol-based cooling system allows for the petal to be powered and tested while cycling, enabling cold tests past $\unit[-35]{\oC}$.

Note that temperatures are measured at the inlet and outlet of the cooling pipes of a petal and are generally lower than the temperature at the sensor due to heat generated by the readout electronics, if they are powered during testing. This was confirmed by a direct temperature measurement using a thermocouple glued directly onto one of the sensors on a petal. These measurements demonstrate a gradient of $\unit[+2]{\oC}$ between the cooling pipe and the surface of the sensors in regions near the cooling pipe and up to a gradient of $\unit[+7]{\oC}$ in regions away from the cooling pipe while the petal is powered.

\subsection{Crack identification}
\label{subsec:cracks}

Cracks are identified using a combination of early breakdown during IV tests, noise results, and visual confirmation. Early breakdown at a voltage smaller than $\unit[-100]{V}$ for a given module is highly indicative of a sensor crack, especially when the early breakdown appears following thermal cycling at a colder temperature. Figure~\ref{fig:worsing_iv_example} shows IV scans of an R0 module tested over many thermal cycles. The worsening nature of the results starting after cycles at $\unit[-60]{\oC}$ is highly indicative of a crack in the sensor.

\begin{figure}[htbp]
    \centering
    \begin{subfigure}[b]{\textwidth}
        \includegraphics[width=\textwidth]{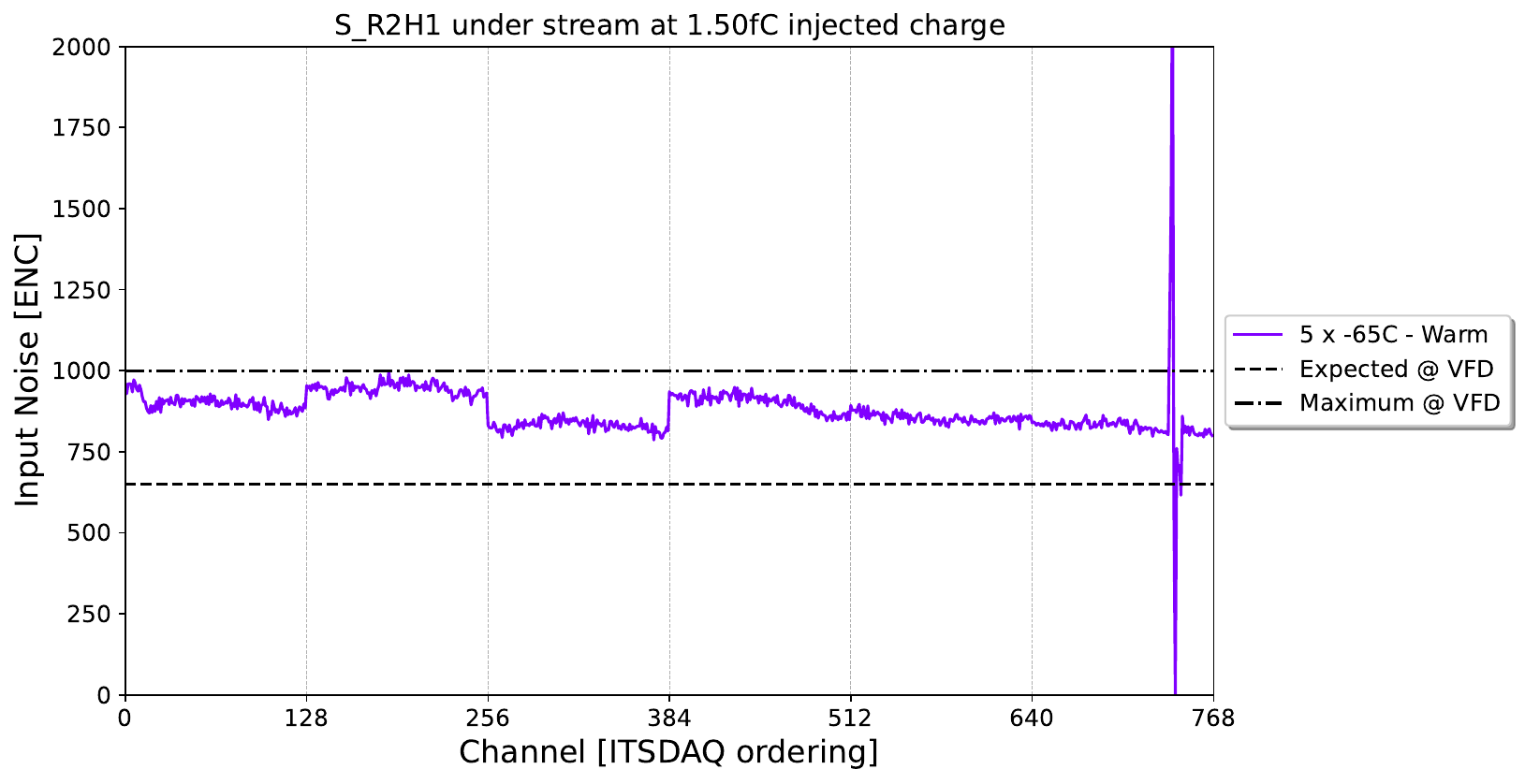}
        \caption{}
        \label{fig:VAN3_S_R2H1_Under_Noise}
    \end{subfigure} \\
    \begin{subfigure}[b]{\textwidth}
        \includegraphics[width=\textwidth]{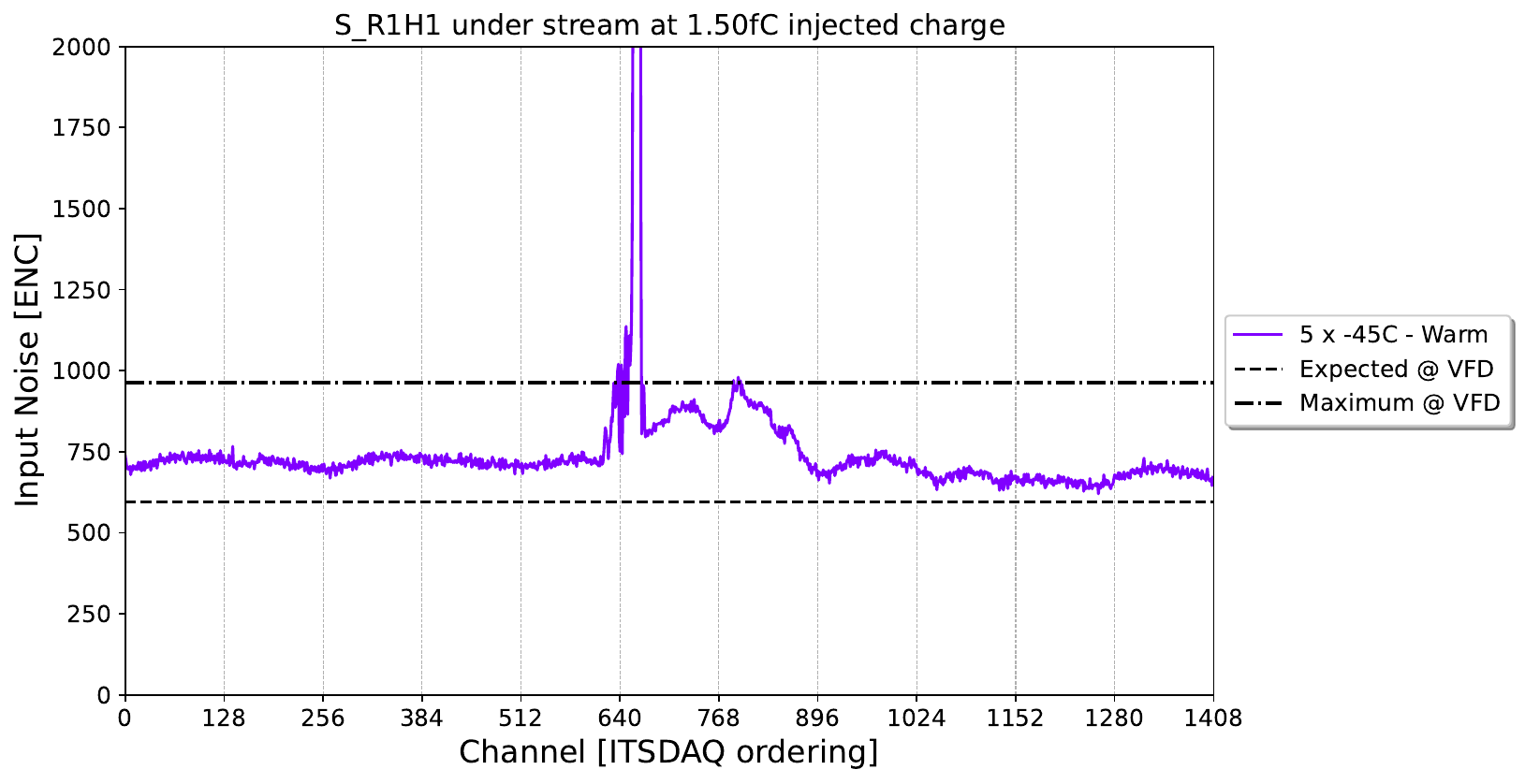}
        \caption{}
        \label{fig:VAN3_S_R1H1_Under_Noise}
    \end{subfigure}
    \caption{(a) R2H1 DAQ scan on a petal (``VAN3'') showing a crack signature around channel~730, characterized by low noise as compared to the neighbouring channels. (b) R1H1 DAQ scan on a petal (``VAN3'') showing a crack signature around channel~640, characterized by higher noise compared to the neighbouring channels.}
    \label{fig:high_low_noise_comp}
\end{figure}

Next, the noise results of the suspect module are investigated, with cracks characterized by a cluster of channels of unusually low or high noise. Figure~\ref{fig:worsing_daq_example} shows DAQ scans for the under stream of an R4 module tested over many thermal cycles. Scans performed before the $\unit[-65]{\oC}$ thermal cycle show a module within specifications, but afterwards the irregular noise results indicate a crack. The overall noise level increase for the $\unit[-65]{\oC}$ scan is the result of a smaller bias voltage due to early breakdown after this cycle. The behaviour of the noise for a cracked module depends on the module geometry and location of the crack. Under stream cracks on sensors with four rows of strips (i.e., R0, R1, and R3 sensors) tend to break the bias resistor off resulting in higher noise, while under stream cracks on sensors with two rows of strips (R2, R4, and R5 sensors) tend to break after the bias resistor resulting in shorter strip lengths and thus lower noise. An example of low and high noise on R2 and R1 modules from the VAN3 petal is shown in figure~\ref{fig:high_low_noise_comp}.

Finally, the module is investigated under a microscope, using the suspect channels as a location guide, to visually confirm a crack. This final step is often impossible to perform without removing the PCBs, and so it is generally done but not always successful.

Since early breakdown and areas with high noise can occur for different reasons, a module is generally required to show at least two of the three criteria associated with cracks before it is considered cracked.

\FloatBarrier
\section{Results from pre-production petals}
\label{sec:results}

This section describes the results obtained from the pre-production petals built to study and mitigate sensor cracking available at the time of writing. For the earliest petals, the default adhesive, SE4445, in the default ``snake-like'' pattern was used for loading (section~\ref{subsec:se4445_petals}). A higher Young's modulus adhesive, Loctite EA 9396 (``HYSOL'')~\cite{hysol:tds}, was studied for some of the petals which followed, deposited in the default snake-like pattern (section~\ref{subsec:hysol_snake}) and later an alternative ``full-coverage'' pattern (section~\ref{subsec:hysol_fc}). Two interposer petals were later assembled, the results of which are not included in this paper. Finally petals relying on the ``Improved Nominal'' solution --- a modification to the PCB-to-sensor adhesive pattern used within the modules and a modification to loading adhesive  pattern --- were assembled (section~\ref{subsec:impnom}). The pre-production petals used in these investigations, along with their loading adhesives and patterns, are listed in table~\ref{tab:listofpetals}. Relevant physical properties for both SE4445 and HYSOL are listed in table~\ref{tab:glue_properties}.

\begin{table}[htbp]
    \centering
    \caption{Pre-production petals used in the cracking mitigation study. The DESY2 petal was loaded using a combination of SE4445 in a snake-like pattern and HYSOL in a full-coverage pattern.}
    \label{tab:listofpetals}
    \begin{tabular}{c | l | l | l}
        \toprule
        Petal                  & Loading adhesive & Loading pattern/feature & Assembly location \\
        \midrule
        VAN1                   & SE4445           & Snake-like              & TRIUMF/SFU \\
        VAN2                   & SE4445           & Snake-like              & TRIUMF/SFU \\
        VAN3                   & HYSOL            & Snake-like              & TRIUMF/SFU \\
        VAN4                   & HYSOL            & Full-coverage           & TRIUMF/SFU \\
        VAN5                   & SE4445           & Interposed modules      & TRIUMF/SFU \\
        VAN6                   & SE4445           & Improved nominal        & TRIUMF/SFU \\
        VAN7                   & SE4445           & Improved nominal        & TRIUMF/SFU \\
        IFIC1                  & SE4445           & Snake-like              & IFIC \\
        IFIC2                  & SE4445           & Snake-like              & IFIC \\
        DESY1                  & SE4445           & Snake-like              & DESY \\
        \multirow{2}{*}{DESY2} & SE4445           & Snake-like              & \multirow{2}{*}{DESY} \\
                               & HYSOL            & Full-coverage           & \\
        DESY3                  & SE4445           & Interposed modules      & DESY \\
        \bottomrule
    \end{tabular}
\end{table}

\begin{table}[htbp]
    \centering
    \caption{A subset of the mechanical, thermal, and electrical properties for the SE4445 and HYSOL adhesives that are relevant to this paper.}
    \label{tab:glue_properties}
    \resizebox{\textwidth}{!}{
        \begin{tabular}{l | l | r | r | r | l}
            \toprule
            Technical name                   & Abbreviated  &        Young's modulus  &    Thermal conductivity  &     Volume resistivity & Radiation \\
                                             & name         & at \unit[25]{\oC} [MPa] & [J/s$\cdot$cm$\cdot$\oC] &    [$\Omega$$\cdot$cm] & hard?     \\
            \midrule
            SE 4445 CV~\cite{se4445:tds}     & SE4445       &   $2.75 \times 10^{+3}$ &    $1.26 \times 10^{-2}$ & $7.00 \times 10^{+15}$ & Yes       \\
            Loctite EA 9396~\cite{hysol:tds} & HYSOL        &      $2 \times 10^{-1}$ &    $2.10 \times 10^{-3}$ & $2.14 \times 10^{+15}$ & Yes       \\
            \bottomrule
        \end{tabular}
    }
\end{table}

In addition to the use of different adhesives and different deposition patterns during loading, many of the pre-production petals were assembled using modules of varying builds due to limitations in component availability at the time. These included a range of pre-production and production hybrids and powerboards as well as different adhesives between the PCBs and the sensors. All of these modifications are reflective of the dynamic nature of the study. Some of these modules were also thermal cycled to $\unit[+40]{\oC}$ during module QC. It was later discovered that at this temperature, modules approached the glass transition temperatures of the adhesives used between the PCBs and the sensors~\cite{Salami:2025nob}, and so the maximum temperature for thermal cycling was lowered to $\unit[+20]{\oC}$. Thus, modules cycled to higher temperatures were more deformed and had a greater tendency to exhibit sensor cracking from thermal stress at low temperatures. Relevant features of the modules on each of the aforementioned pre-production petals are summarized in table~\ref{tab:module_builds}.

\begin{table}[htbp]
    \centering
    \caption{Module builds for pre-production petals. For each module, the adhesive used at the PCB-to-sensor interface is shown; some modules were assembled using the Epolite FH-5313 epoxy (``Polaris'')~\cite{polaris:tds}. Also shown is the maximum temperature each module was thermal cycled (TC) to during module QC. In two cases, the adhesive (DESY1: R3) and temperature (IFIC1: R2) were not recorded (``Unknown''). Polaris was the original prototyping adhesive which was subsequently discontinued by the manufacturer. Eccobond F112 and AA Bond F112 were later used as they have similar mechanical properties. SE4445 was used for modules assembled to investigate noise features not relevant for the scope of this paper~\cite{Dyckes:2024bhv} and is not comparable with the other adhesives because it is much softer.}
    \label{tab:module_builds}
    \resizebox{0.65\textwidth}{!}{
        \begin{tabular}{c | c | l r | l r}
            \toprule
            & & \multicolumn{2}{c|}{\textbf{Main side}} & \multicolumn{2}{c}{\textbf{Secondary side}} \\
            Petal & Module & Adhesive & TC [$\oC$] & Adhesive & TC [$\oC$] \\
            \midrule
            \multirow{6}{*}{VAN1}  & R5 & Polaris        & +40 & Polaris        & +40 \\
                                   & R4 & Polaris        & +40 & Polaris        & +40 \\
                                   & R3 & Polaris        & +40 & Polaris        & +40 \\
                                   & R2 & Eccobond F112  & +40 & SE4445         & +40 \\
                                   & R1 & Polaris        & +40 & AA Bond F112   & +40 \\
                                   & R0 & Polaris        & +40 & Polaris        & +20 \\
            \midrule
            \multirow{6}{*}{VAN2}  & R5 & Eccobond F112  & +20 & AA Bond F112   & +20 \\
                                   & R4 & Eccobond F112  & +20 & AA Bond F112   & +20 \\
                                   & R3 & AA Bond F112   & +20 & Eccobond F112  & +20 \\
                                   & R2 & Polaris        & +20 & AA Bond F112   & +20 \\
                                   & R1 & Eccobond F112  & +20 & AA Bond F112   & +40 \\
                                   & R0 & AA Bond F112   & +20 & Eccobond F112  & +20 \\
            \midrule
            \multirow{6}{*}{VAN3}  & R5 & Eccobond F112  & +20 & AA Bond F112   & +20 \\
                                   & R4 & Eccobond F112  & +20 & Polaris        & +40 \\
                                   & R3 & AA Bond F112   & +20 & Eccobond F112  & +20 \\
                                   & R2 & Eccobond F112  & +20 & Eccobond F112  & +20 \\
                                   & R1 & Polaris        & +40 & Polaris        & +40 \\
                                   & R0 & AA Bond F112   & +20 & AA Bond F112   & +20 \\
            \midrule
                                   & R5 & Eccobond F112  & +20 & Eccobond F112  & +20 \\
                                   & R4 & Eccobond F112  & +20 & Eccobond F112  & +20 \\
            VAN4/5/6/7             & R3 & Eccobond F112  & +20 & Eccobond F112  & +20 \\
            and DESY2/3            & R2 & Eccobond F112  & +20 & Eccobond F112  & +20 \\
                                   & R1 & Eccobond F112  & +20 & Eccobond F112  & +20 \\
                                   & R0 & Eccobond F112  & +20 & Eccobond F112  & +20 \\
            \midrule
            \multirow{6}{*}{IFIC1} & R5 & Polaris        & +40 & Polaris        & +20 \\
                                   & R4 & Polaris        & +40 & Polaris        & +40 \\
                                   & R3 & AA Bond F112   & +20 & Eccobond F112  & +20 \\
                                   & R2 & Polaris        & +20 & AA Bond F112   & Unknown \\
                                   & R1 & SE4445         & +40 & Eccobond F112  & +40 \\
                                   & R0 & Eccobond F112  & +40 & AA Bond F112   & +20 \\
            \midrule
            \multirow{6}{*}{IFIC2} & R5 & Eccobond F112  & +20 & Eccobond F112  & +20 \\
                                   & R4 & AA Bond F112   & +20 & Eccobond F112  & +20 \\
                                   & R3 & AA Bond F112   & +20 & Eccobond F112  & +20 \\
                                   & R2 & Eccobond F112  & +20 & AA Bond F112   & +20 \\
                                   & R1 & Eccobond F112  & +20 & AA Bond F112   & +20 \\
                                   & R0 & Eccobond F112  & +20 & AA Bond F112   & +20 \\
            \midrule
            \multirow{6}{*}{DESY1} & R5 & AA Bond F112   & +20 & Eccobond F112  & +20 \\
                                   & R4 & Eccobond F112  & +20 & AA Bond F112   & +40 \\
                                   & R3 & AA Bond F112   & +20 & Unknown        & +40 \\
                                   & R2 & Eccobond F112  & +40 & AA Bond F112   & +20 \\
                                   & R1 & Polaris        & +40 & AA Bond F112   & +20 \\
                                   & R0 & Polaris        & +40 & Polaris        & +20 \\
            \bottomrule
        \end{tabular}
    }
\end{table}

The VAN1 and IFIC1 petals were installed in a larger test structure~\cite{Arling:2024tty} before cracking was discovered and therefore could not be tested beyond $\unit[-35]{\oC}$. In contrast, early staves showed a number of cracks when tested at $\unit[-35]{\oC}$, highlighting an intrinsic difference in the thermo-mechanical performance between petals and staves~\cite{Tishelman-Charny:2024wxu}. The VAN5 and DESY3 petals were assembled using interposed modules, and therefore their results are beyond the scope of this paper.

Note that throughout this section, modules and the cracks which occur on modules are labelled with a prefix ``M'' or ``S'' denoting the main and secondary sides, respectively, and a suffix denoting the module (or split module).\footnote{For example, ``M\_R0'' corresponds to the main side R0 module and ``S\_R3M0'' corresponds to the secondary side R3M0 module.} This is done for brevity.
\FloatBarrier

\subsection{SE4445 default petals: VAN2, IFIC2, and DESY1}
\label{subsec:se4445_petals}

\begin{figure}[htbp]
    \centering
    \includegraphics[width=0.95\textwidth]{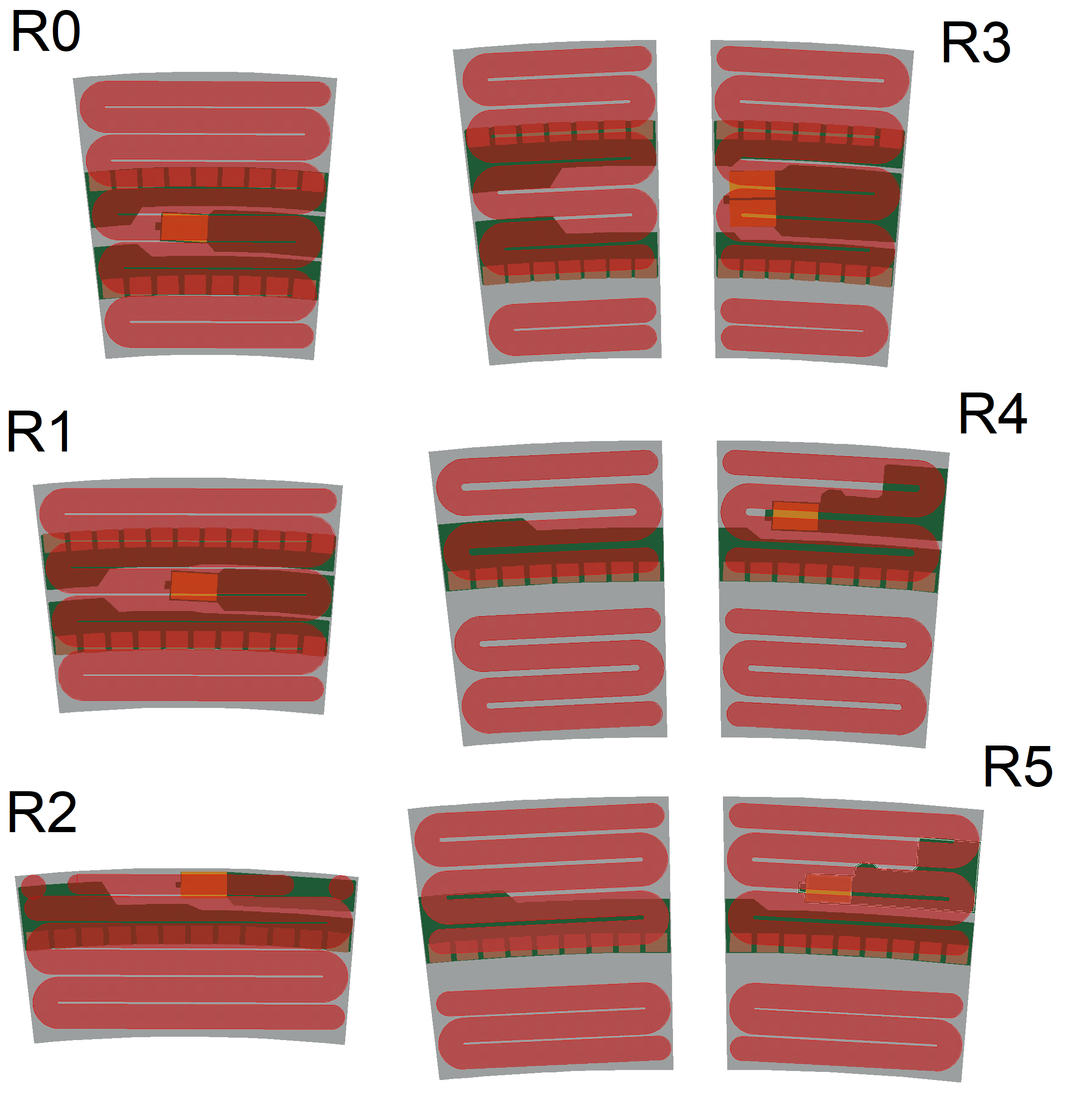}
    \caption{SE4445 loading adhesive in a snake-like pattern as seen for each type of module. The patterns are shown in red and the PCBs are shown in green. The different module types are not necessarily to scale.}
    \label{fig:default_snake_all}
\end{figure}

The initial petals built using the default adhesive, SE4445, in the default snake-like pattern (i.e., with no cracking mitigation strategies applied) are referred to as the default petals; these include the VAN2, IFIC2, and DESY1 petals. They were loaded with modules from the first iteration of pre-production and therefore used different PCB-to-sensor adhesives and  were thermal cycled to different temperatures as part of module QC. They were loaded using SE4445 in a snake-like pattern as shown in figure~\ref{fig:default_snake_all}. This default loading pattern was developed to maximize coverage under each module.

\subsubsection{VAN2}
\label{subsubsec:van2_result}

\begin{figure}[htbp]
    \begin{minipage}[b]{\linewidth}
        \centering
        \includegraphics[width=\textwidth]{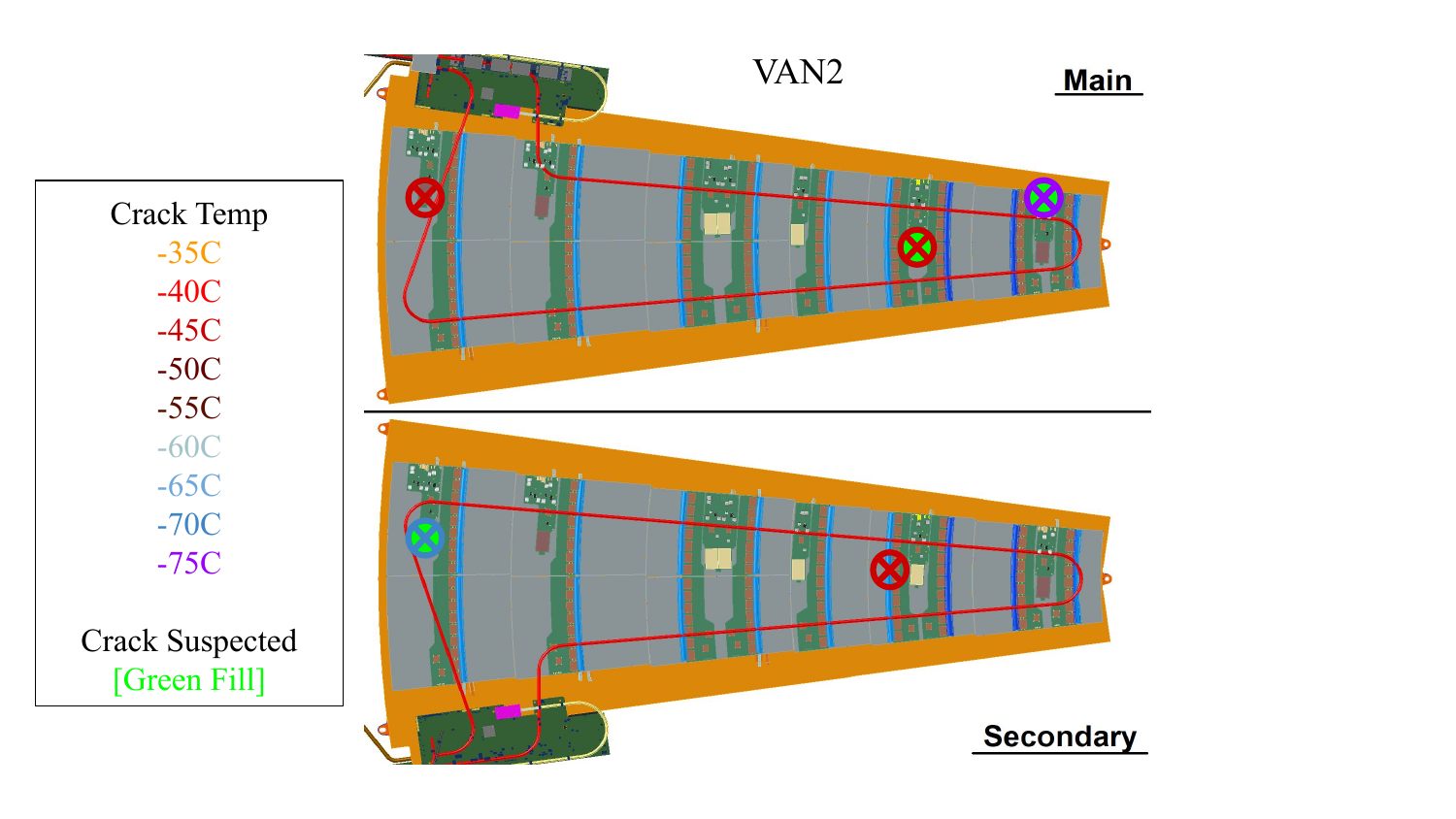}
    \end{minipage}
    \begin{minipage}[b]{\linewidth}
        \centering
        \begin{tabular}{ c | c | c || c | c | c}
            \toprule
            \shortstack{\textbf{Temperature} \\ \textbf{[$\oC$]}}         &
            \shortstack{\textbf{Module}      \\ \textbf{crack}}           &
            \shortstack{\textbf{Visually}    \\ \textbf{confirmed [Y/N]}} &
            \shortstack{\textbf{Temperature} \\ \textbf{[$\oC$]}}         &
            \shortstack{\textbf{Module}      \\ \textbf{crack}}           &
            \shortstack{\textbf{Visually}    \\ \textbf{confirmed [Y/N]}} \\
            \midrule
            $-45$ & M\_R5M0 & Y &
            $-70$ & S\_R5M0 & N \\ \midrule
            $-45$ & M\_R1   & Y &
            $-75$ & M\_R0   & N \\ \midrule
            $-45$ & S\_R1   & N &
                  &         &   \\ \bottomrule
        \end{tabular}
    \end{minipage}
    \caption{Suspected and confirmed crack locations on the VAN2 petal, after each crack signature was seen in the IV and/or DAQ scans.}
    \label{fig:van2_overview}
\end{figure}

The VAN2 petal was initially subjected to passive thermal cycling down to $\unit[-45]{\oC}$, after which it was transferred to a setup capable of active cycling to much colder temperatures, once the first cracks were identified. The petal was then actively thermally cycled to $\unit[-75]{\oC}$. An overview of the suspected and visually confirmed crack locations is presented in figure~\ref{fig:van2_overview}.

The S\_R1 crack was located between two ABCs on the R1H1 hybrid. The M\_R5M0 crack was confirmed visually by removing the hybrid and was found to originate at a small adhesive patch directly under the powerboard. The remaining cracks were not confirmed visually: their locations are inferred from their noise signatures and from the confirmed crack locations on future petals (section~\ref{subsec:hysol_snake}).

\subsubsection{IFIC2}
\label{subsubsec:ific2_result}

\begin{figure}[htbp]
    \begin{minipage}[b]{\linewidth}
        \centering
        \includegraphics[width=\textwidth]{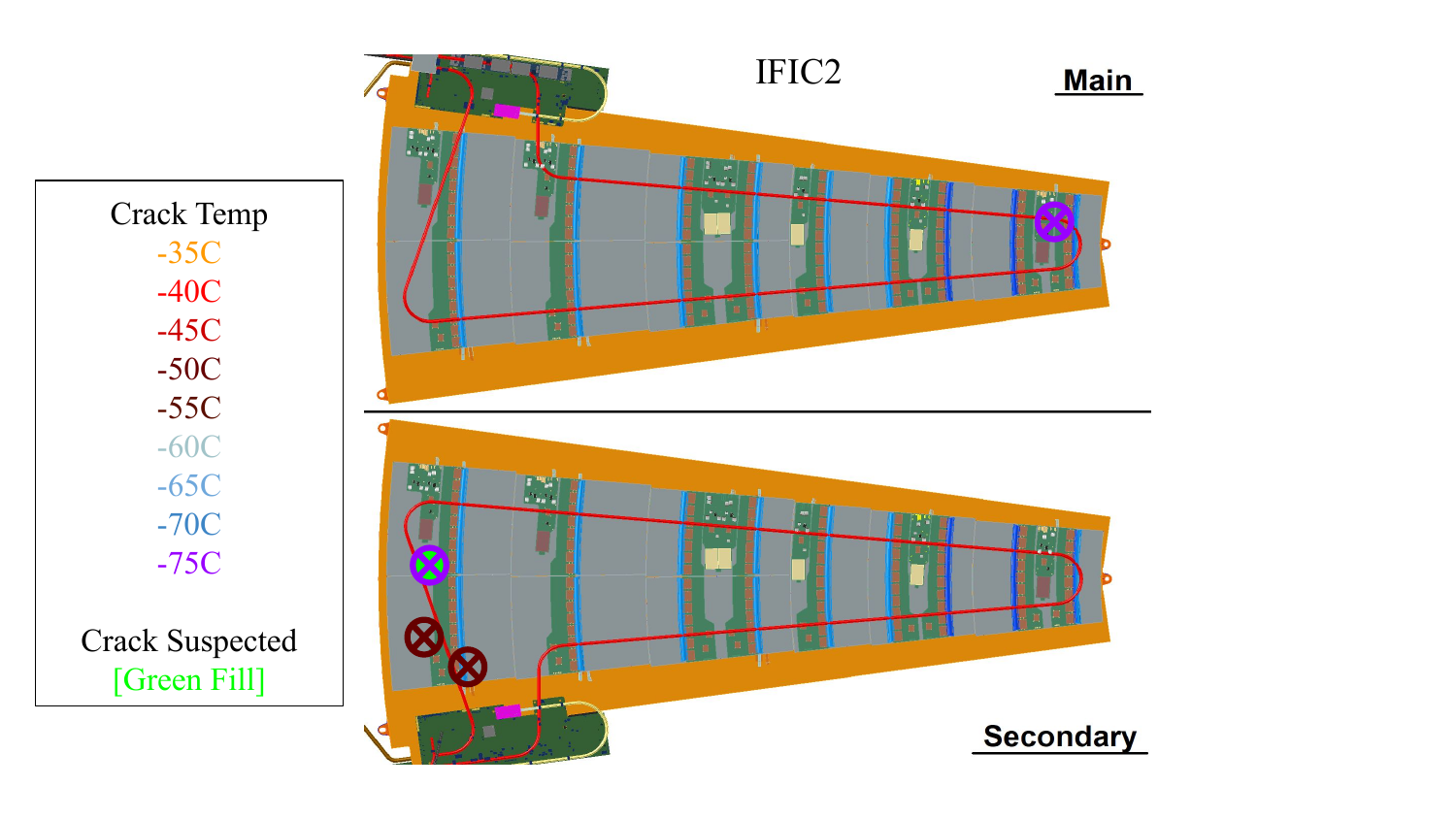}
    \end{minipage}
    \begin{minipage}[b]{\linewidth}
        \centering
        \begin{tabular}{ c | c | c || c | c | c}
            \toprule
            \shortstack{\textbf{Temperature} \\ \textbf{[$\oC$]}}         &
            \shortstack{\textbf{Module}      \\ \textbf{crack}}           &
            \shortstack{\textbf{Visually}    \\ \textbf{confirmed [Y/N]}} &
            \shortstack{\textbf{Temperature} \\ \textbf{[$\oC$]}}         &
            \shortstack{\textbf{Module}      \\ \textbf{crack}}           &
            \shortstack{\textbf{Visually}    \\ \textbf{confirmed [Y/N]}} \\
            \midrule
            $-50$\dag  & S\_R5M1 & Y &
            $-75$      & S\_R5M0 & N \\ \midrule
            $-50$\ddag & S\_R5M1 & Y &
            $-75$      & M\_R0   & N \\ \bottomrule
        \end{tabular}
    \end{minipage}
    \caption{Suspected and confirmed crack locations on the IFIC2 petal, after each crack signature was seen in the IV and/or DAQ scans. (\dag{}) The S\_R5M1 module cracked after first thermal cycle. (\ddag{}) The S\_R5M1 module cracked after second thermal cycle.}
    \label{fig:ific2_overview}
\end{figure}

The IFIC2 petal was actively thermal cycled and tested down to $\unit[-75]{\oC}$. It was later discovered that the main side R4 module had a defective powerboard and could not be powered, while the secondary side R4 module had very high noise for reasons not relevant for this discussion. Therefore, data collected from these modules were excluded from further analysis. An overview of the suspected and visually confirmed crack locations is presented in figure~\ref{fig:ific2_overview}.

\begin{figure}[htbp]
    \centering
    \includegraphics[width=\textwidth]{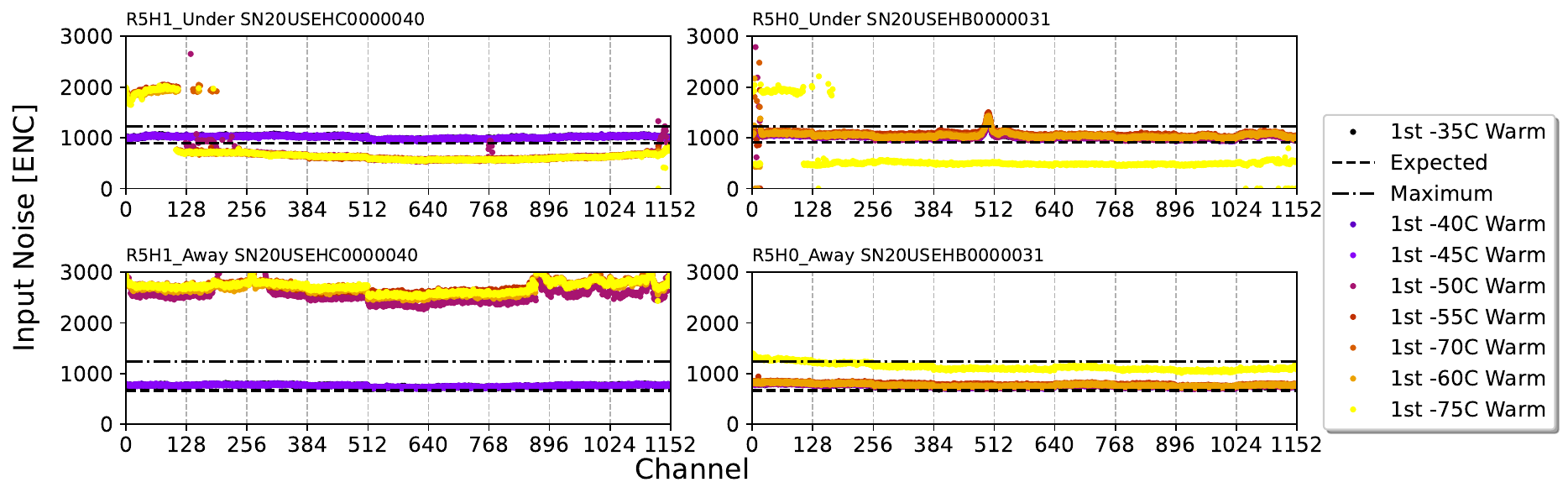}
    \caption{Noise results for the secondary side R5 module on the IFIC2 petal after the first cycle at each temperature. Under stream cracks across both sensors show lower noise than expected.}
    \label{fig:ifi2_S_R5_extreme}
\end{figure}

\begin{figure}[htbp]
    \centering
    \includegraphics[width=0.9\textwidth]{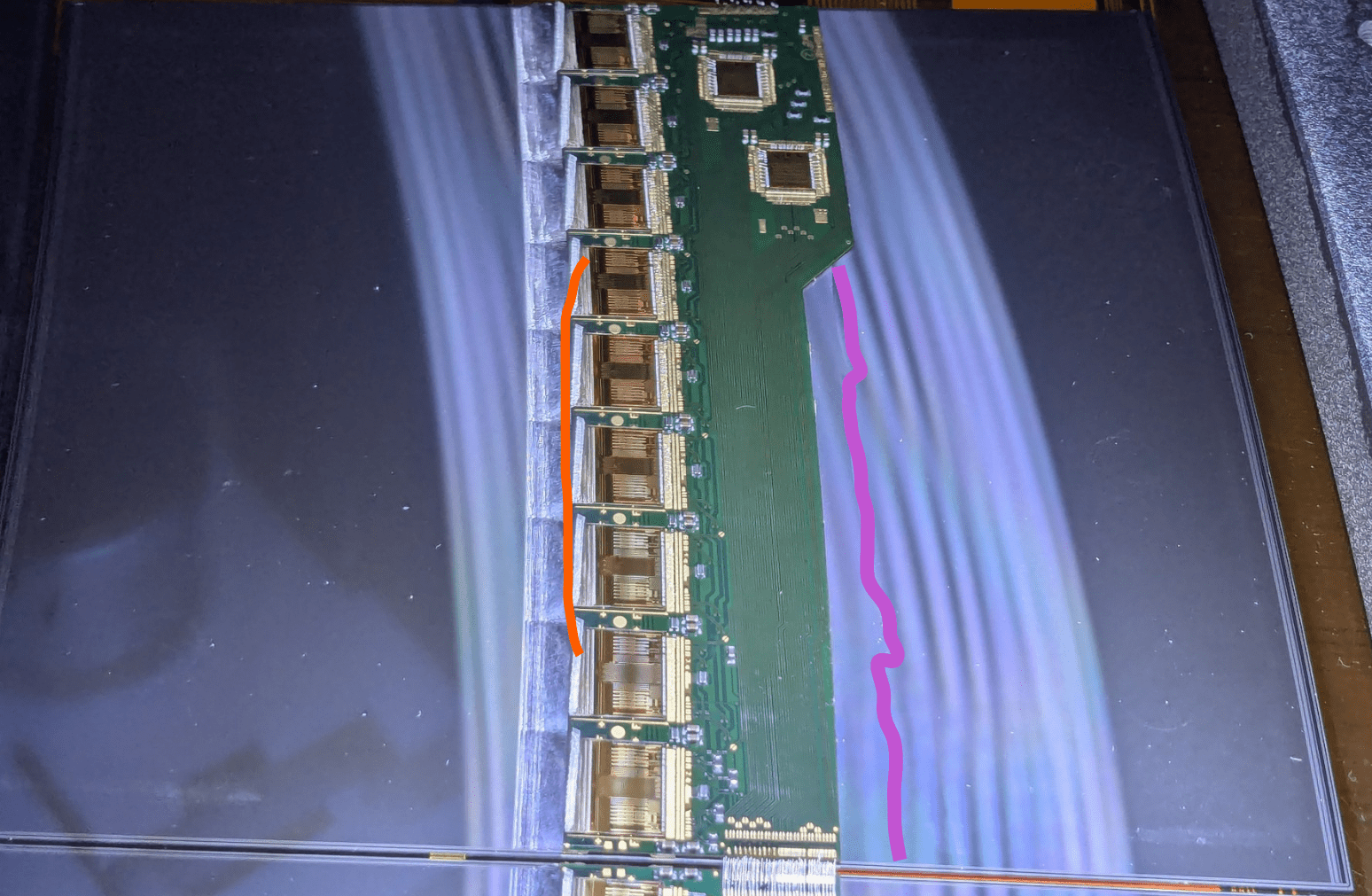}
    \caption{Visually confirmed crack locations on secondary side R5M1 module on the IFIC2 petal after the $\unit[-50]{\oC}$ cycles. The crack indicated in orange is a typical hairline crack, while crack indicated in purple fully separates the sensor into two pieces.}
    \label{fig:ific2_extreme_visual}
\end{figure}

The S\_R5M1 cracks were visually confirmed after the first and second $\unit[-50]{\oC}$ cycles, extending nearly the width of the entire module. Two additional crack signatures were detected after the $\unit[-75]{\oC}$ cycle on the S\_R5M0 and M\_R0 modules. Figure~\ref{fig:ifi2_S_R5_extreme} presents the noise for the S\_R5 module, arranged to match the module geometry, after the first cycle at each temperature. Similar cracks in the under streams of both sensors indicates similar internal stresses. Both cracks were visually confirmed to be above the hybrids. Figure~\ref{fig:ific2_extreme_visual} presents the two cracks seen on the S\_R5M1 module: the crack indicated in orange on the left-hand side, propagating under the wire bonds to ABCs, was identified after the first $\unit[-50]{\oC}$ cycle and corresponds to a typical hairline crack. The crack indicated in purple on the right-hand side was observed after the second $\unit[-50]{\oC}$ cycle; this crack is noteworthy as it splits the sensor into two physical pieces, which is highly unusual.

\subsubsection{DESY1}
\label{subsubsec:desy1_result}

\begin{figure}[htbp]
    \begin{minipage}[b]{\linewidth}
        \centering
        \includegraphics[width=\textwidth]{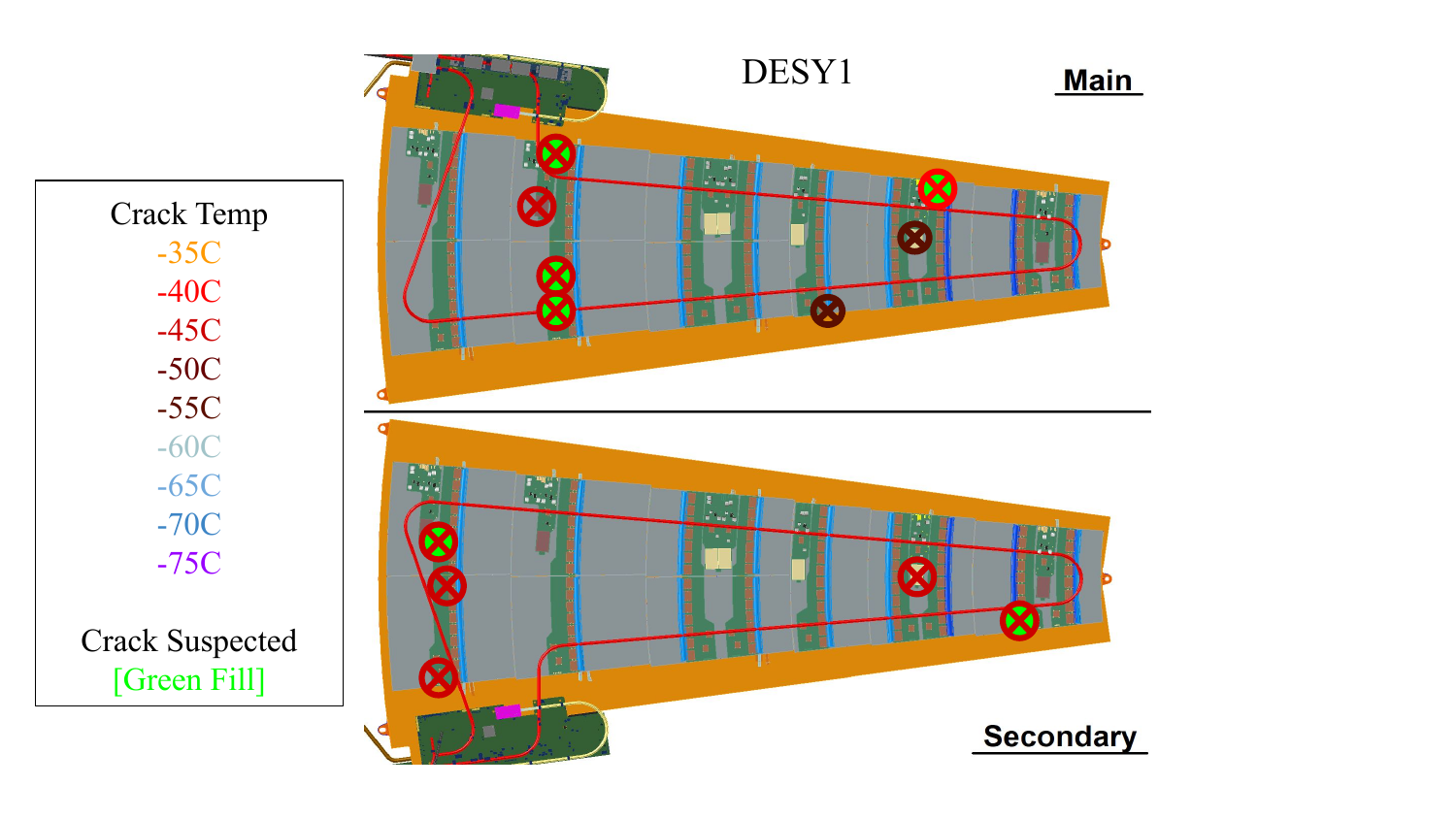}
    \end{minipage}
    \begin{minipage}[b]{\linewidth}
        \centering
        \begin{tabular}{ c | c | c || c | c | c}
            \toprule
            \shortstack{\textbf{Temperature} \\ \textbf{[$\oC$]}}         &
            \shortstack{\textbf{Module}      \\ \textbf{crack}}           &
            \shortstack{\textbf{Visually}    \\ \textbf{confirmed [Y/N]}} &
            \shortstack{\textbf{Temperature} \\ \textbf{[$\oC$]}}         &
            \shortstack{\textbf{Module}      \\ \textbf{crack}}           &
            \shortstack{\textbf{Visually}    \\ \textbf{confirmed [Y/N]}} \\
            \midrule
            $-40$ & M\_R1   & N &
            $-45$ & S\_R5M0 & N \\ \midrule
            $-45$ & M\_R4M0 & Y\dag &
            $-45$ & S\_R5M1 & Y\ddag \\ \midrule
            $-45$ & M\_R4M1 & N &
            $-55$ & M\_R1   & Y \\ \midrule
            $-45$ & S\_R0   & N &
            $-55$ & M\_R2   & Y \\ \midrule
            $-45$ & S\_R1   & Y &
                  &         &   \\ \bottomrule
        \end{tabular}
    \end{minipage}
    \caption{Suspected and confirmed crack locations on the DESY1 petal, after crack signature is seen in DAQ and/or IV scan. (\dag) The M\_R4M0 module had two crack signatures of which the one under the DC-DC converter was visually confirmed. (\ddag) The S\_R5M1 module had two crack signatures of which both were visually confirmed. }
    \label{fig:desy1_overview}
\end{figure}

The DESY1 petal was actively thermal cycled and tested down to $\unit[-55]{\oC}$. An overview of the visually confirmed and suspected crack locations on DESY1 are presented in figure~\ref{fig:desy1_overview}. For the M\_R4 and S\_R5 modules, multiple crack signatures were observed. One of these was visually confirmed on the S\_R5 module, located along the edge of the module and crossing the HV connection to the petal core.

The S\_R1 and M\_R4 cracks were both visually confirmed near the DC-DC converter. The M\_R2 crack was visually identified beneath the leftmost ABC, while the second M\_R1 crack was similarly seen propagating from beneath the DC-DC converter.

This petal exhibited a larger number of cracks at comparatively warmer temperatures than the two petals assembled with a similar build. Further investigation revealed that many of the modules loaded on this petal had been assembled with a modified PCB-to-sensor adhesive pattern. This modification was later determined to increase mechanical stress and contributed to the development of the Improved Nominal petal design (see section~\ref{subsec:impnom}).

\subsubsection{Lessons learned from SE4445 default petals}
\label{subsubsec:se4445_default_conclusion}

While the SE4445 default petals did not exhibit any cracking at $\unit[-35]{\oC}$, one petal showed cracking after thermal cycling at $\unit[-45]{\oC}$, a temperature which modules can expect to experience at some time during the operation of the ITk. Consequently, the design of these petals was considered insufficient to mitigate sensor cracking within the relevant QC temperature range. However, it is worth noting that modules on petals generally exhibited fewer cracks than modules on staves at the same temperatures, suggesting reduced mechanical stress. The mechanism for this is not completely understood. but it is suspected to be related to the use of curved PCBs in end-cap modules. For example, it is possible that the curved PCBs in end-cap modules result in less overlap with the lattice directions of the silicon sensor than the straight PCBs in barrel modules.

Furthermore, most of the cracked modules on the default petals did not conform to final production specifications (e.g., using Eccobond F112 for assembly), which may also affect the number of cracks observed. Among these tested prototypes, the IFIC2 petal included the largest number of modules built to specification, with the first crack occurring after thermal cycling at $\unit[-50]{\oC}$.

\FloatBarrier

\subsection{HYSOL default petal: VAN3}
\label{subsec:hysol_snake}

\begin{figure}[htbp]
    \centering
    \begin{subfigure}{0.47\textwidth}
        \includegraphics[width=\textwidth]{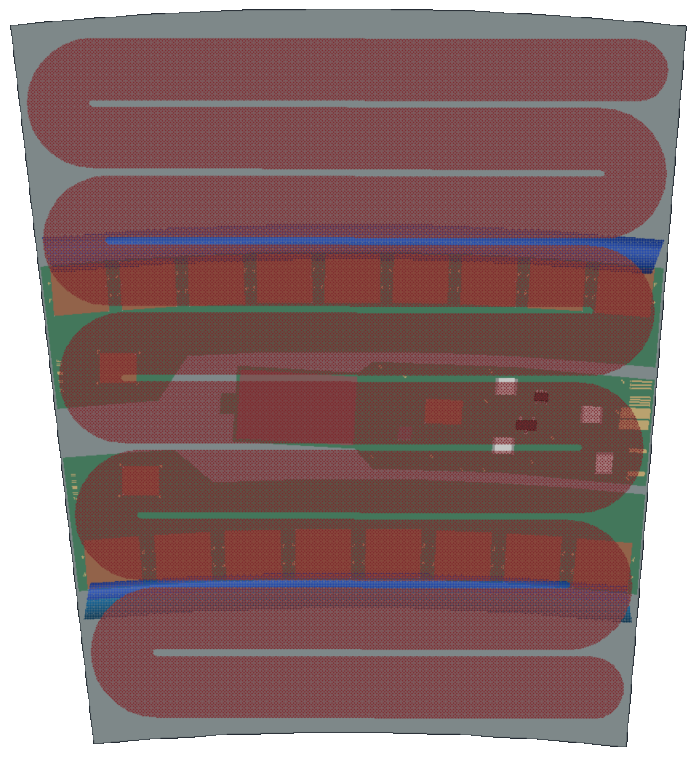}
        \caption{}
    \end{subfigure}
    \begin{subfigure}{0.46\textwidth}
        \includegraphics[width=\textwidth]{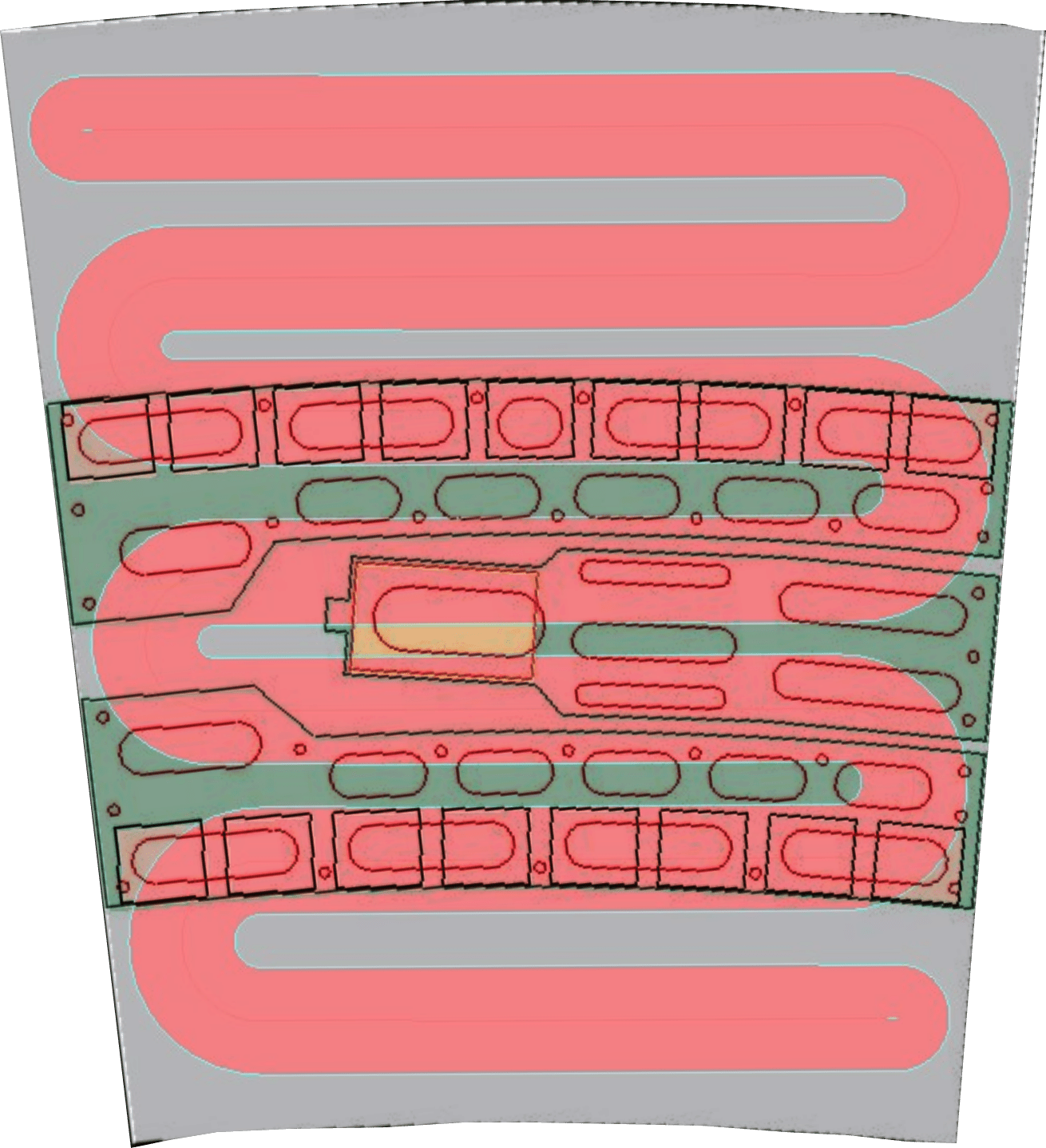}
        \caption{}
    \end{subfigure}
    \caption{(a) Default pattern for an R0 module loaded with SE4445 in a snake-like pattern. (b) Default pattern for an R0 module loaded with HYSOL in a snake-like pattern overlaid with the hybrids, powerboard, and PCB-to-sensor adhesive pattern; the PCB-to-sensor adhesive pattern in (a) is the same as that shown in (b). HYSOL supports the PCB edges and gaps. The PCB-to-sensor adhesive pattern provides support to ASICs and DC-DC converter where small ``glue-dots'' are used to bridge the areas between adhesive patches.}
    \label{fig:se445_hysol_comparison}
\end{figure}

The HYSOL mitigation strategy was based largely on following the mitigation HYSOL staves built for the barrel and simulations showing a reduction in stress at crack locations when using a harder (i.e., higher Young's modulus) adhesive such as HYSOL. Since the use of a higher modulus adhesive changed the stress distribution between different layers, replacing the adhesive required an accompanying adjustment of the loading pattern under each module. The recommendations for the modified loading pattern were made based on the barrel module geometry:\footnote{Individual end-cap module geometries were only later simulated due to time constraints.}
\begin{itemize}
    \item Full HYSOL coverage in the gap(s) between the hybrid(s) and the powerboard;
    \item Coverage of the long edges of the PCB-to-sensor adhesive patterns to at least \unit[2]{mm} into the adhesive layer;
    \item Extending the HYSOL as far as possible towards the sensor edges;
    \item Coverage of the gaps between the hybrids on R0, R1, and R3 modules as much as possible.
\end{itemize}
The new loading patterns developed for this petal based on these recommendations are seen was additionally optimized to provide support to the DC-DC converter and the ASICs. Figure~\ref{fig:se445_hysol_comparison} shows a comparison of the loading patterns used in the SE4445 and HYSOL default petals for an R0 module.

The VAN3 petal was loaded with modules from various stages of module pre-production, as indicated in table~\ref{tab:module_builds}. The main side R4 and R5 modules were known to have issues related to their powerboards, resulting in higher current draw and higher noise. No electrical tests of these modules were performed after the initial IV and DAQ scans. The VAN3 petal was passively thermal cycled and tested down to $\unit[-70]{\oC}$.

With the exception of M\_R2 and M\_R3M0 modules, no cracks were visually observed; this led to the assumption that an unusually large fraction of cracks formed under the PCBs. The M\_R2 crack was discovered after the $\unit[-40]{\oC}$ cycle, at the edge of the module under the wire bonds on the leftmost ABC. The M\_R3M0 crack signature appeared after the $\unit[-45]{\oC}$ cycle, between the powerboard and the R3H2 hybrid near the DC-DC converter.

\begin{figure}[htbp]
    \centering
    \includegraphics[width=0.4\textwidth]{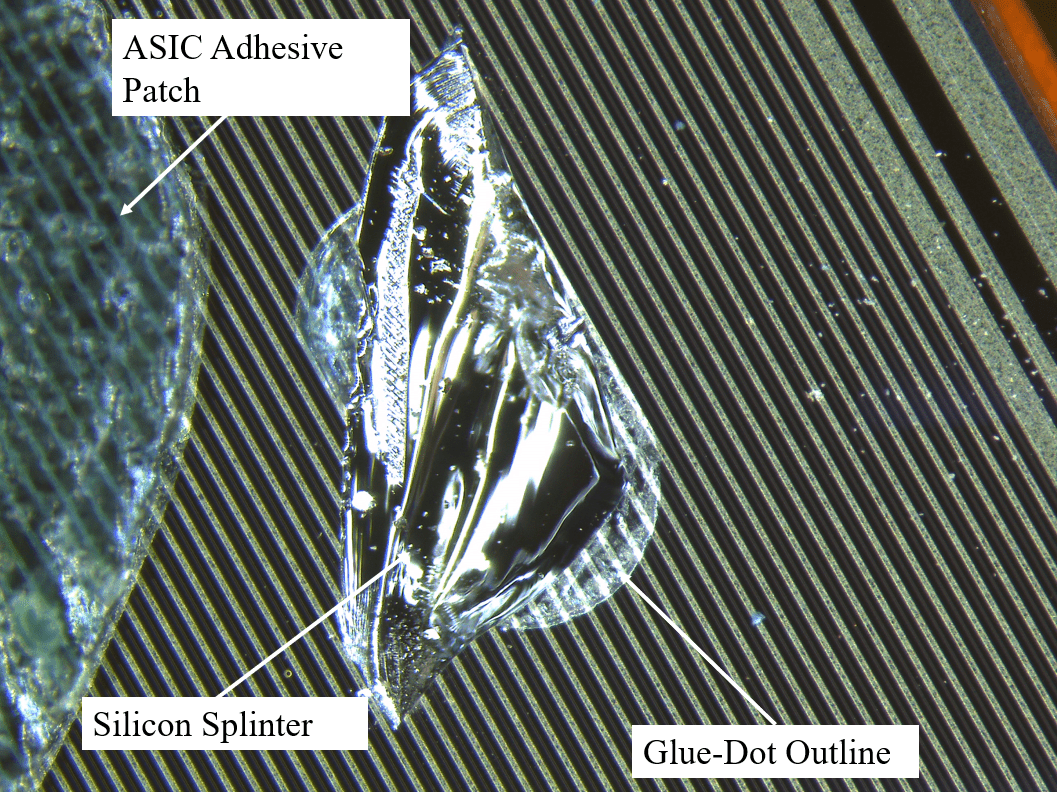}
    \includegraphics[width=0.59\textwidth]{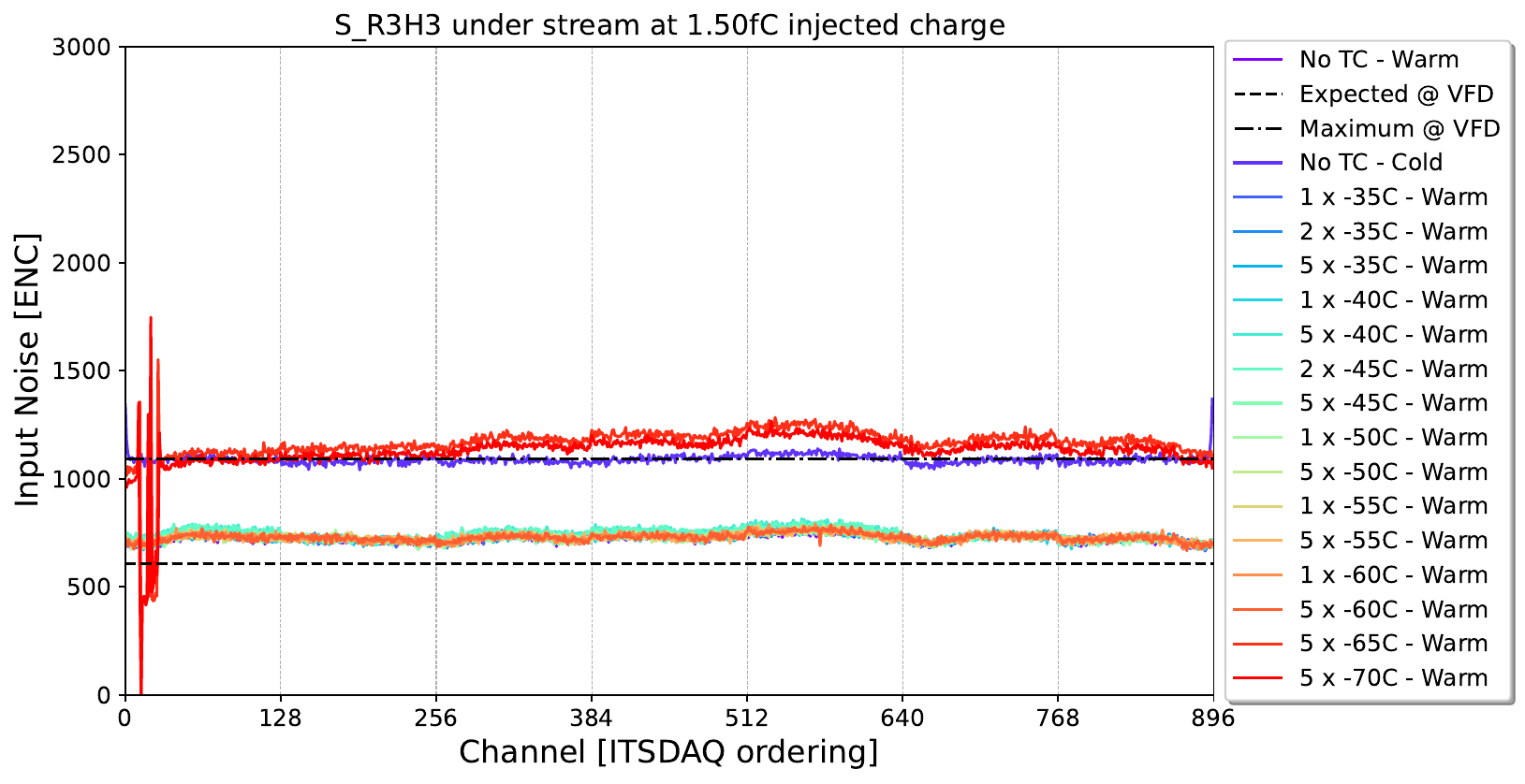} \\
    \includegraphics[width=0.4\textwidth]{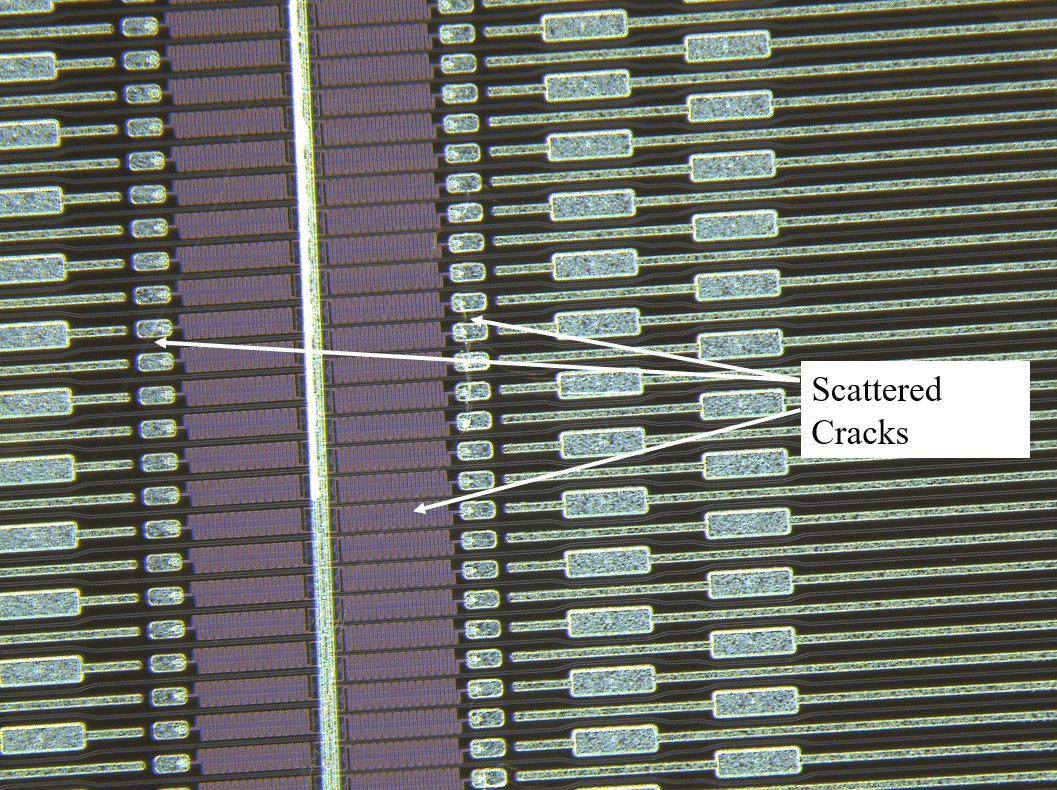}
    \includegraphics[width=0.59\textwidth]{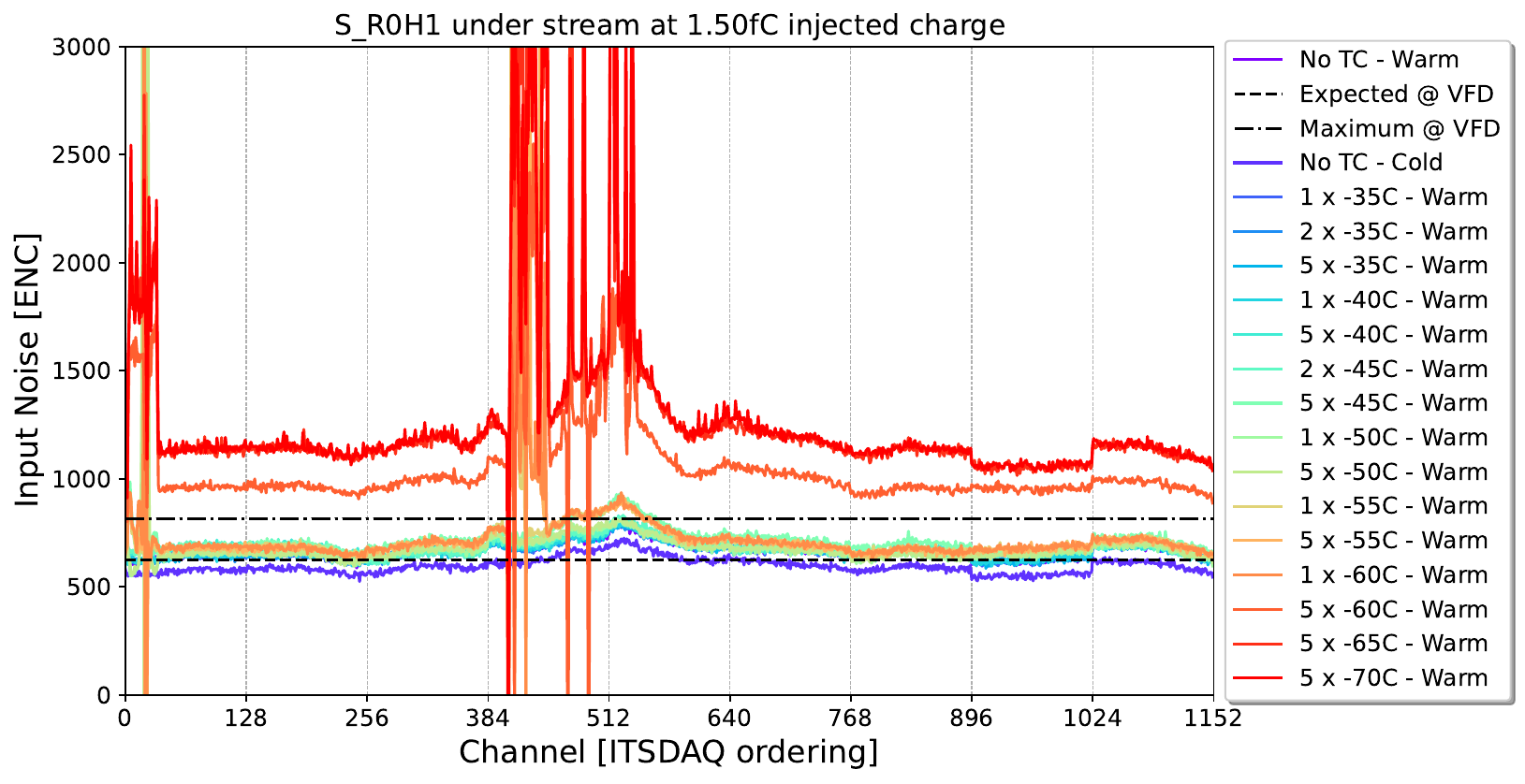} \\
    \includegraphics[width=0.4\textwidth]{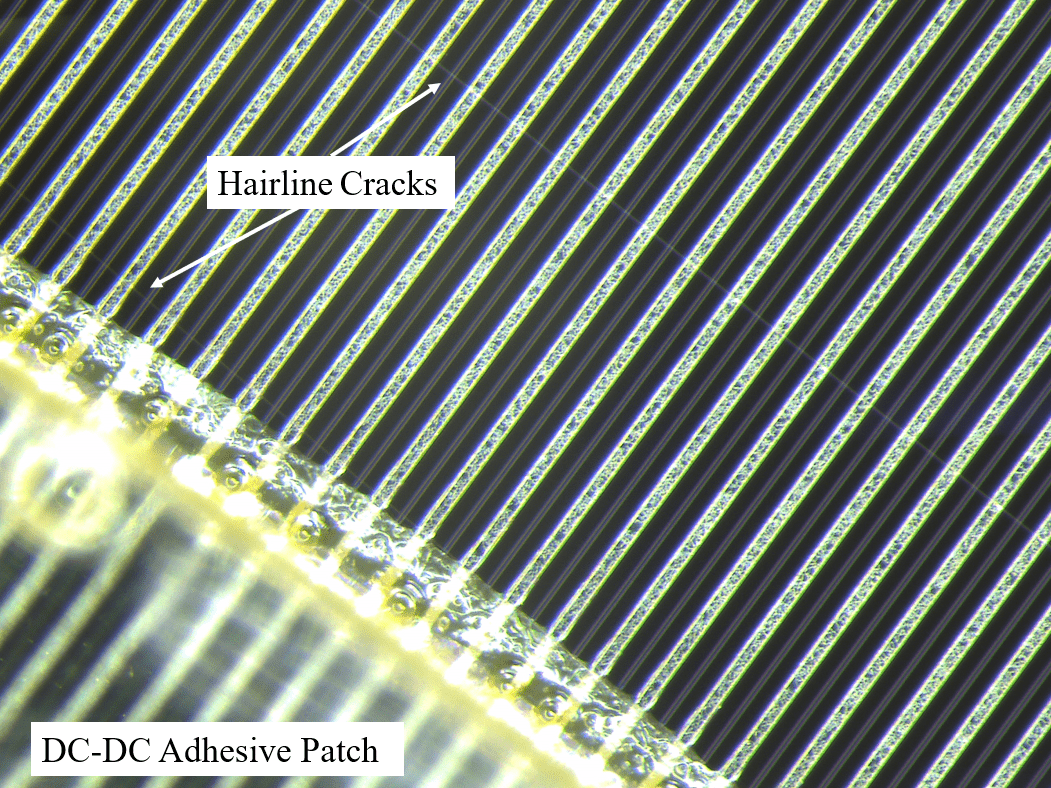}
    \includegraphics[width=0.59\textwidth]{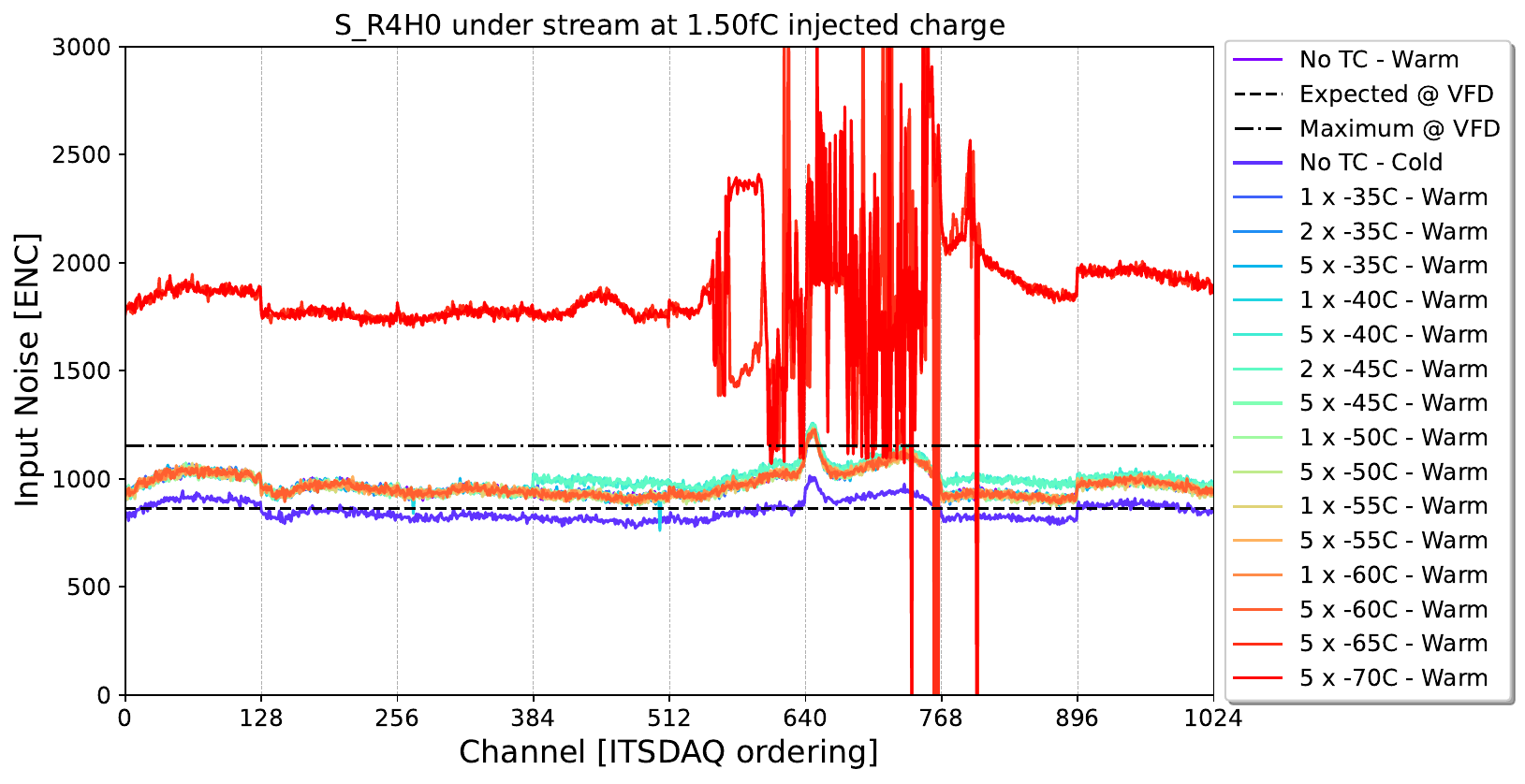} \\
    \caption{Three categories of cracks discovered from the VAN3 petal. Each example is from the secondary side. (Top) Removed silicon splinter at an edge glue-dot on the R3M1 module. This type of crack has a localized noise shape which does not grow with colder thermal cycles. (Middle) Multiple short cracks across bias resistors on the R0 module. This type of crack has high and low noise channels spanning one or two ABCs. (Bottom) Single long crack on the R3M1 module spanning three ABCs. This type of crack is often seen near the adhesive outline under the DC-DC converter.}
    \label{fig:hysol_snake_crack_examples}
\end{figure}

\begin{figure}[htbp]
    \begin{minipage}[b]{\linewidth}
        \centering
        \includegraphics[width=\textwidth]{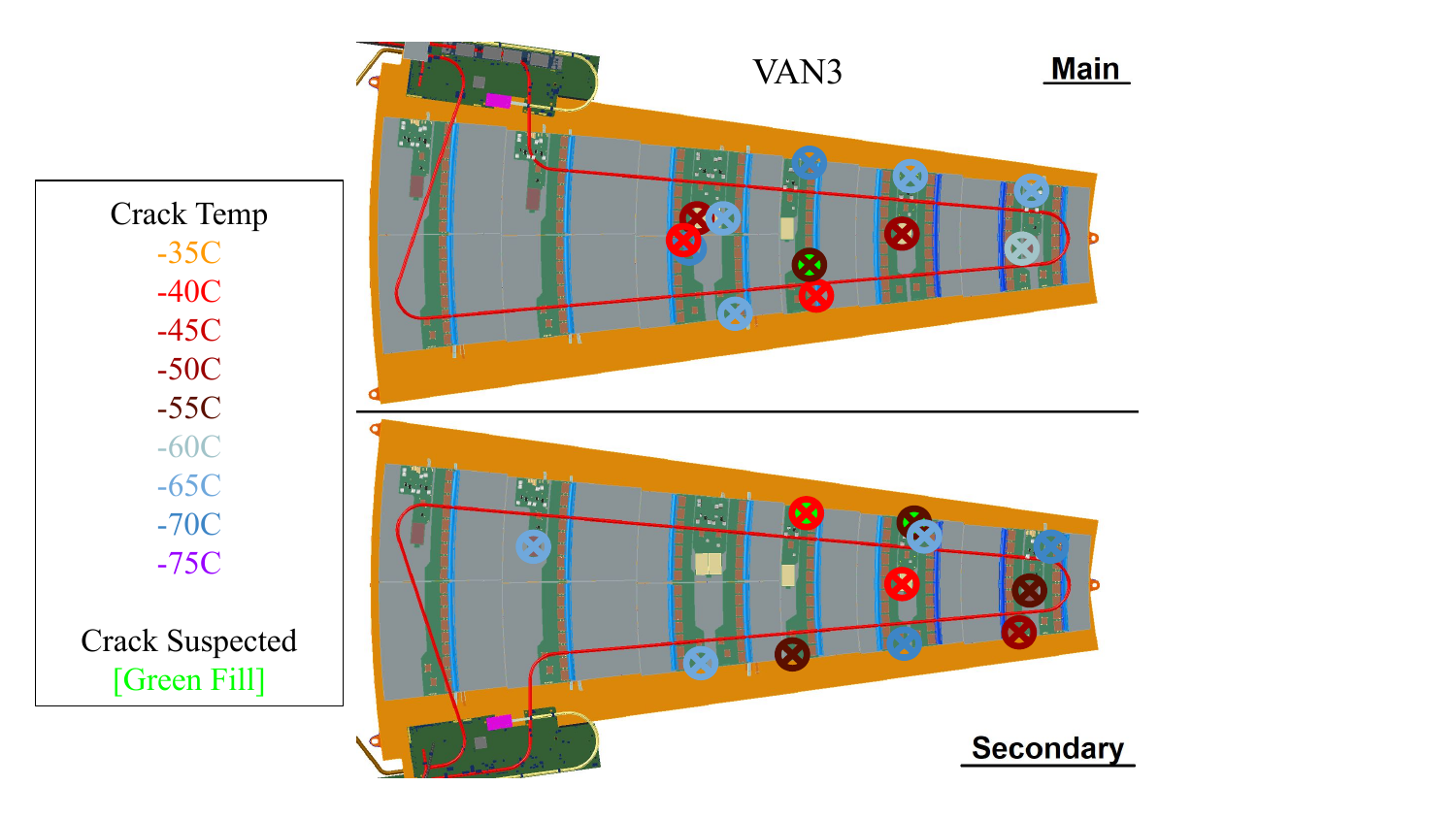}
    \end{minipage}
    \begin{minipage}[b]{\linewidth}
        \centering
        \begin{tabular}{ c | c | c || c | c | c}
            \toprule
            \shortstack{\textbf{Temperature} \\ \textbf{[$\oC$]}}         &
            \shortstack{\textbf{Module}      \\ \textbf{crack}}           &
            \shortstack{\textbf{Visually}    \\ \textbf{confirmed [Y/N]}} &
            \shortstack{\textbf{Temperature} \\ \textbf{[$\oC$]}}         &
            \shortstack{\textbf{Module}      \\ \textbf{crack}}           &
            \shortstack{\textbf{Visually}    \\ \textbf{confirmed [Y/N]}} \\
            \midrule
            $-40$ & S\_R1   & Y &
            $-65$ & M\_R0   & Y \\ \midrule
            $-40$ & M\_R3M1 & Y &
            $-65$ & M\_R1   & Y \\ \midrule
            $-40$ & M\_R2   & Y &
            $-65$ & M\_R3M0 & Y \\ \midrule
            $-40$ & S\_R2   & N &
            $-65$ & M\_R3M1 & Y \\ \midrule
            $-50$ & M\_R1   & Y &
            $-65$ & S\_R1   & Y \\ \midrule
            $-50$ & M\_R3M0 & Y &
            $-65$ & S\_R3M1 & Y \\ \midrule
            $-50$ & S\_R0   & Y &
            $-65$ & S\_R4M0 & Y \\ \midrule
            $-55$ & M\_R2   & N &
            $-70$ & M\_R2   & Y \\ \midrule
            $-55$ & S\_R0   & Y &
            $-70$ & M\_R3M1 & Y \\ \midrule
            $-55$ & S\_R1   & N &
            $-70$ & S\_R0   & Y \\ \midrule
            $-55$ & S\_R2   & Y &
            $-70$ & S\_R1   & Y \\ \midrule
            $-60$ & M\_R0   & Y &
                  &         &  \\ \bottomrule
        \end{tabular}
    \end{minipage}
    \caption{Suspected and confirmed crack locations on the VAN3 petal, after each crack signature was seen in the IV and/or DAQ scans.}
    \label{fig:van3_overview}
\end{figure}

To investigate the remaining crack signatures after completing the testing program at $\unit[-70]{\oC}$, the PCBs of modules with early breakdown were removed after heating the module adhesive past its melting point, ensuring that the removal of the PCBs would not further damage the sensors. Visual inspection of suspected regions under the PCBs yielded three categories of cracks. Figure~\ref{fig:hysol_snake_crack_examples} shows an image of the different types of crack found along with the noise signatures. An overview of the suspected and confirmed crack locations are shown in figure~\ref{fig:van3_overview}. These include twenty confirmed cracks and three additional suspected cracks that were not visually confirmed due to excessive adhesive coverage on the sensor surface.

The first category of cracks was where silicon splinters detached from the sensors together with the PCBs. These types of cracks were always found near glue-dots, small deposits of adhesive in the PCB-to-sensor pattern that are used to bridge the areas between adhesive patches. Specifically, ten cracks were found near glue-dots at the edges of the modules that had localized noise signatures which did not grow with successive thermal cycling. These types of cracks have minimal or absent HYSOL support underneath. These splinters suggested that the glue-dots without HYSOL support create a high stress location that cracks but is not allowed to propagate due to the glue-dot outline. The thickness of the silicon splinters (including the glue) also varied between cracks, ranging from \unit[70--400]{\um}, indicating that this type of cracking is not uniform.

The second category of cracks discovered are multiple short and scattered cracks across the bias resistors under the hybrids or powerboards, depending on the module. These types of cracks had noise signatures that range across multiple ABCs and grow with successive thermal cycling. These cracks occur in the gaps between the hybrids and powerboards on the edge of the HYSOL pattern. It is believed that minimal HYSOL support underneath creates multiple high stress locations that crack at various temperatures.

The final category of cracks discovered were single long cracks spanning an area of two to three ABCs. Specifically, six cracks originated from and propagated along the nearby hybrid-to-sensor adhesive. These cracks occurred under the DC-DC converter and were on the edge of the HYSOL support underneath the sensor. The noise signature for these cracks also propagated across multiple ABCs.

\subsubsection{Lessons learned from HYSOL default petal}
\label{subsubsec:hysol_snake_conclusion}

\begin{figure}[htbp]
    \centering
    \begin{subfigure}{0.47\textwidth}
        \includegraphics[width=\textwidth]{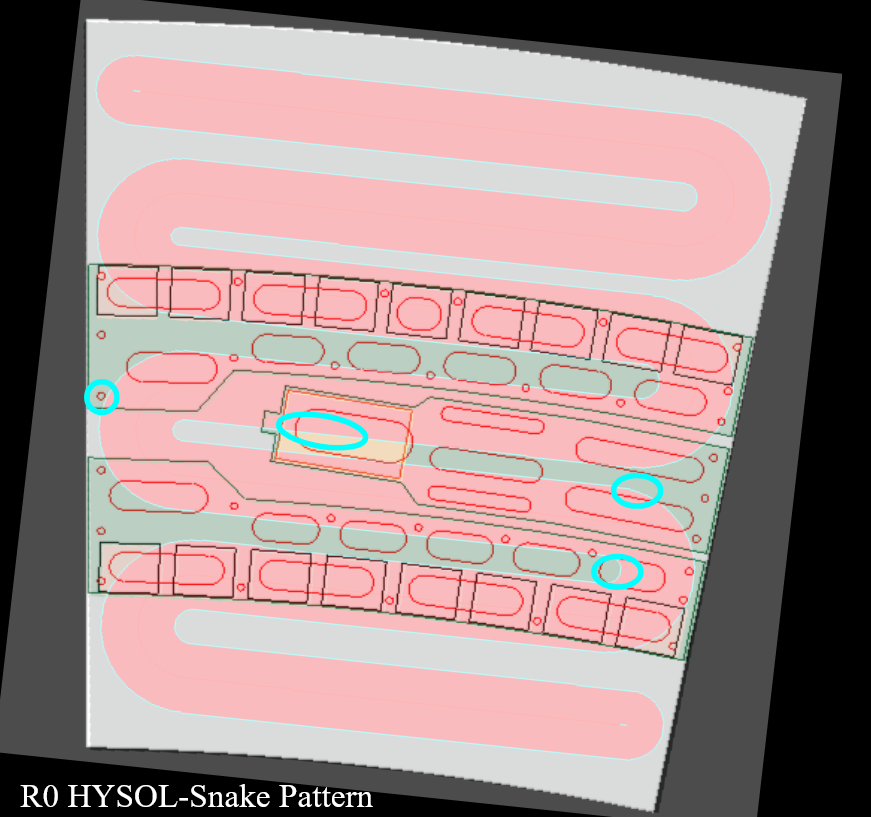}
        \caption{}
    \end{subfigure}
    \begin{subfigure}{0.97\textwidth}
        \includegraphics[width=\textwidth]{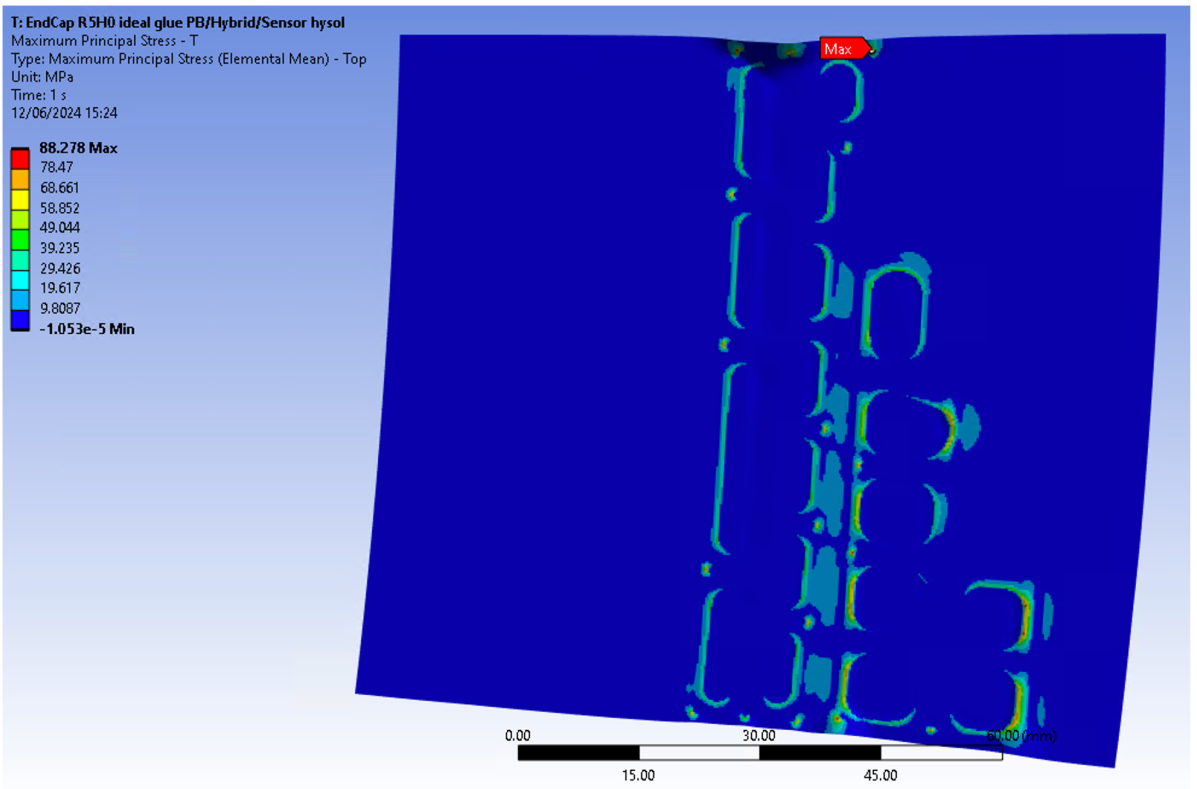}
        \caption{}
    \end{subfigure}
    \caption{(a) R0 loading pattern (solid red area) from figure~\ref{fig:se445_hysol_comparison} overlaid with regions where cracks were discovered (cyan ellipses). (b) Stress simulation of an R5 module using HYSOL in a snake-like loading pattern and a PCB-to-sensor pattern including glue-dots, showing local stress maxima where small glue-dots between PCB and sensor coincide with insufficient HYSOL support underneath.}
    \label{fig:hysol_snake_crack_locations}
\end{figure}

The HYSOL mitigation strategy was based on the premise that a harder adhesive would reduce the internal stress; however, in the VAN3 petal, a larger number of cracks (as compared to SE4445 default petals) were discovered and included a previously unobserved type of crack: silicon splinters. The combined result of the loading and PCB-to-sensor adhesive patterns led to regions of gaps with little-to-no HYSOL support and regions along the module edges with glue-dots and little-to-no HYSOL support. The cracks found on the VAN3 petal largely coincided with these regions and occurred mostly in different regions from the SE4445 petals, underneath PCBs and primarily along their short edges. After the VAN3 petal had finished thermal cycling, simulations of individual end-cap module geometries were implemented to confirm the suspected connection between the glue-dots in the PCB-to-sensor pattern and concentrated stress on the sensor. The simulation confirmed these adhesive areas to be those with the highest stress, as shown in figure~\ref{fig:hysol_snake_crack_locations}.

\FloatBarrier

\subsection{HYSOL optimized petal: VAN4}
\label{subsec:hysol_fc}

\begin{figure}[htbp]
    \centering
    \includegraphics[width=0.9\textwidth]{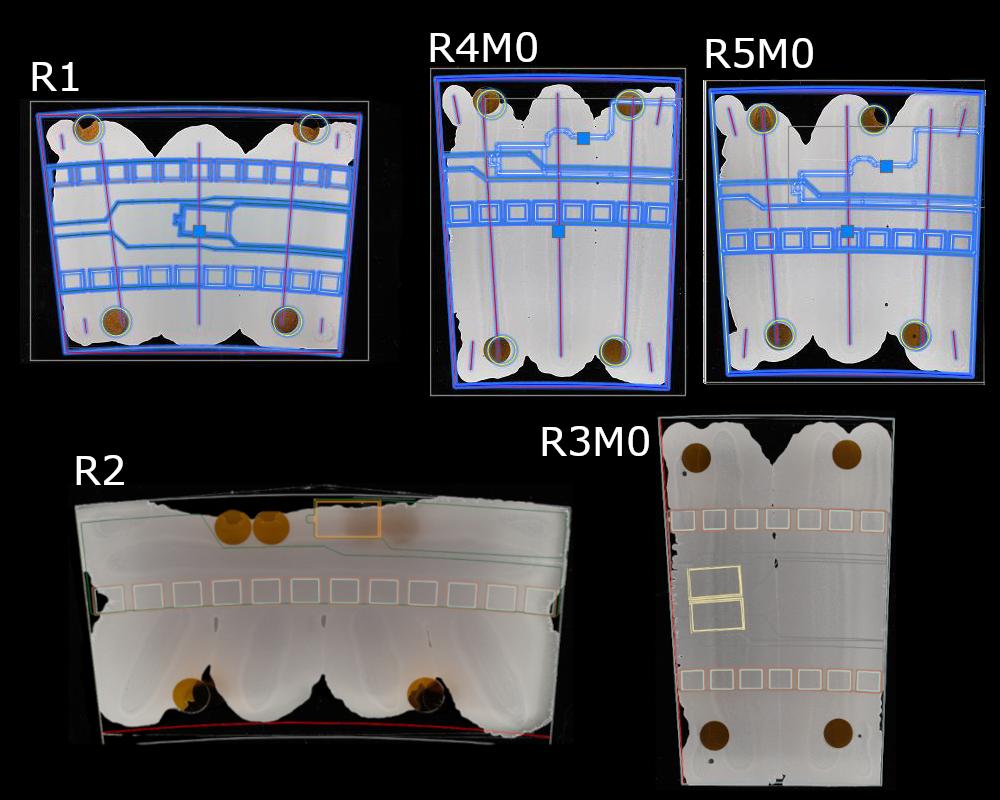}
    \caption{Examples of mechanical glass sensors loaded with HYSOL, shown in grey, in a full-coverage pattern, i.e.\ a pattern covering the entire area under each PCB. PCB patterns for each module type are additionally overlaid. The full-coverage patterns for R0, R3M1, R4M1, and R5M1 modules are similar to those shown above. The different module types are not necessarily to scale.}
    \label{fig:hysol_fc_pattern}
\end{figure}

Following the results of the VAN3 petal (section~\ref{subsec:hysol_snake}) and the confirmation from simulations that glue-dots induce stress, the VAN4 petal was loaded using HYSOL in an optimized or ``full-coverage'' pattern, as shown in figure~\ref{fig:hysol_fc_pattern}. ``Full-coverage'' refers to covering the entire area between all PCBs as well as several millimetres beyond the PCB edges by the HYSOL loading pattern. While the results of the VAN3 petal indicated that modules without glue-dots were beneficial for the petal, a shortage of components for the assembly of new modules resulted in only modules with the detrimental PCB-to-sensor pattern being available for the VAN4 petal. It was therefore assembled with the expectation of silicon splinters in the corresponding module regions, based on previous experience.

\begin{figure}[htbp]
    \centering
    \includegraphics[width=0.97\textwidth]{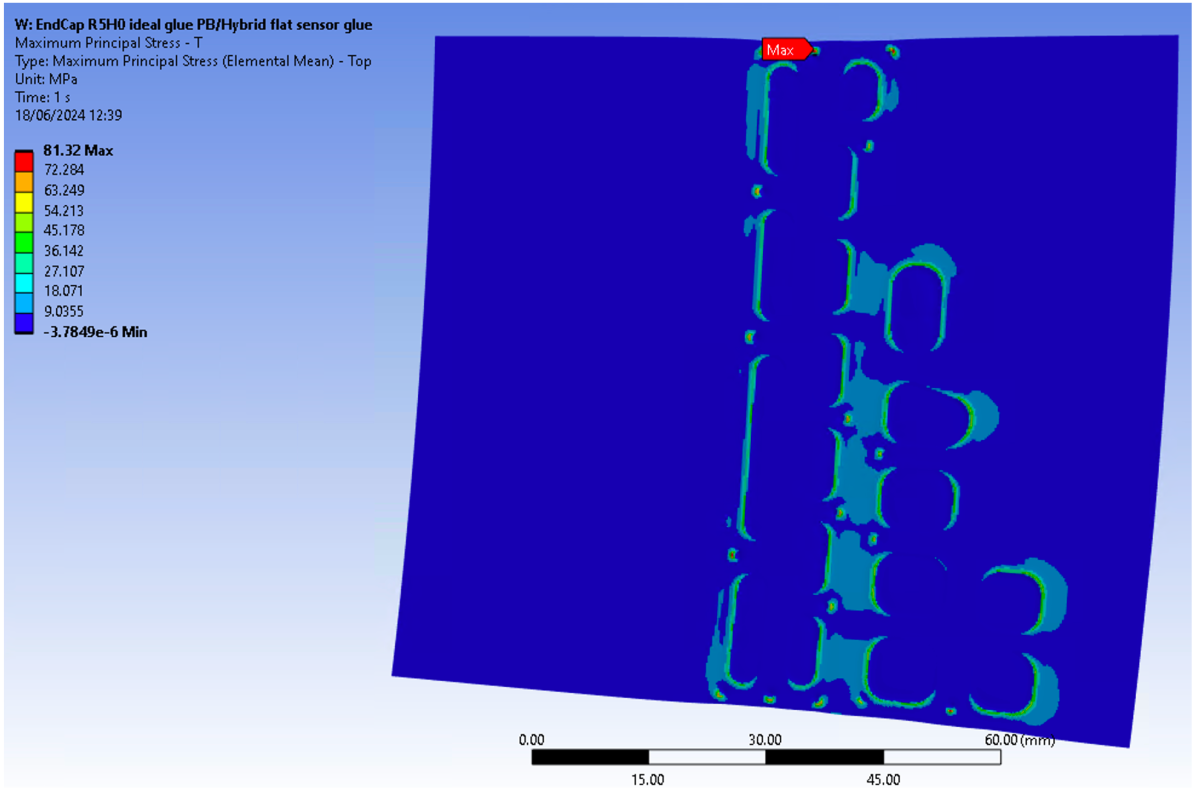}
    \caption{Stress simulation of an R5 module using HYSOL in a full-coverage loading pattern and a PCB-to-sensor pattern including glue-dots. The regions of highest stress occur at glue-dots and in particular along the edges, with the rest of the module showing lower stress compared to the snake-like pattern.}
    \label{fig:hysol_fc_with_dots_sim}
\end{figure}

The goal of the VAN4 petal was to address the cracks present in the gaps between hybrids and powerboard of the snake-like loading pattern. Figure~\ref{fig:hysol_fc_with_dots_sim} shows the stress simulation for an R5 module with HYSOL in a full-coverage pattern and modules with glue-dots, as used on the VAN4 petal.

\begin{figure}[htbp]
    \begin{minipage}[b]{\linewidth}
        \centering
        \includegraphics[width=\textwidth]{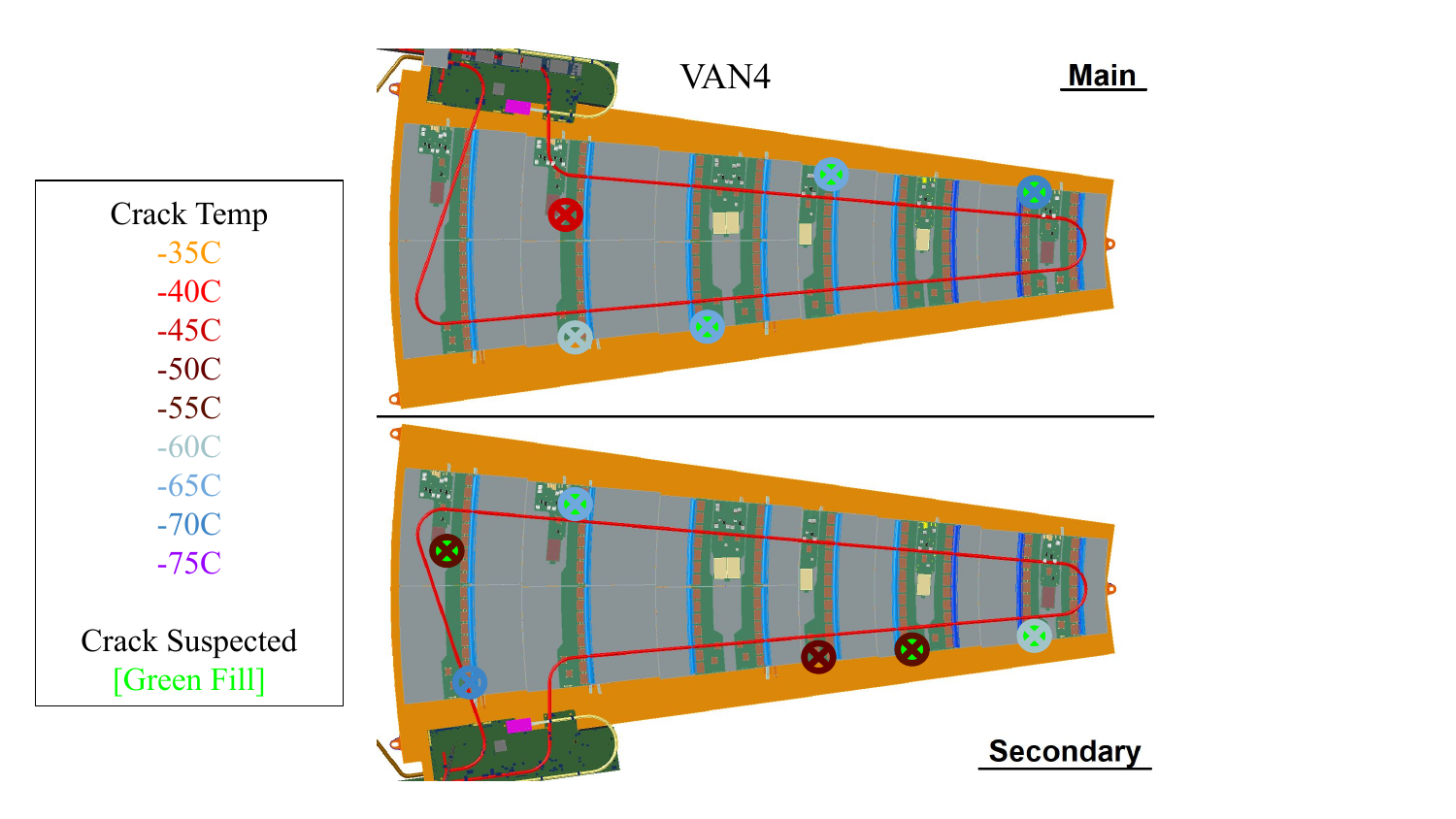}
    \end{minipage}
    \begin{minipage}[b]{\linewidth}
        \centering
        \begin{tabular}{ c | c | c || c | c | c}
            \toprule
            \shortstack{\textbf{Temperature} \\ \textbf{[$\oC$]}}         &
            \shortstack{\textbf{Module}      \\ \textbf{crack}}           &
            \shortstack{\textbf{Visually}    \\ \textbf{confirmed [Y/N]}} &
            \shortstack{\textbf{Temperature} \\ \textbf{[$\oC$]}}         &
            \shortstack{\textbf{Module}      \\ \textbf{crack}}           &
            \shortstack{\textbf{Visually}    \\ \textbf{confirmed [Y/N]}} \\
            \midrule
            $-45$ & M\_R4M0 & Y &
            $-65$ & M\_R2   & N \\ \midrule
            $-50$ & S\_R2   & Y &
            $-65$ & S\_R4M0 & N \\ \midrule
            $-55$ & S\_R1   & N &
            $-65$ & M\_R3M1 & N \\ \midrule
            $-55$ & S\_R5M0 & N &
            $-70$ & M\_R0   & N \\ \midrule
            $-60$ & S\_R0   & N &
            $-70$ & S\_R5M1 & Y \\ \midrule
            $-60$ & M\_R4M1 & Y &
                  &         &  \\ \bottomrule
        \end{tabular}
    \end{minipage}
    \caption{Suspected and confirmed crack locations on the VAN4 petal, after each crack signature was seen in the IV and/or DAQ scans.}
    \label{fig:van4_overview}
\end{figure}

Figure~\ref{fig:van4_overview} presents an overview of the confirmed and suspected cracks on the VAN4 petal. By the end of the thermal cycling schedule, there were eleven suspected crack locations --- a significant reduction compared to the VAN3 petal --- none of them visible. Ten of these were at or near glue-dots and one was under the DC-DC converter of the M\_R4M0 module; the latter crack appeared after the $\unit[-45]{\oC}$ cycle. In order to investigate the crack locations and patterns, PCBs were removed for an investigation of the silicon underneath. In order to continue petal tests afterwards, only the hybrids and powerboards on the M\_R4, S\_R2, S\_R4, and S\_R5 modules were removed to check for crack locations.

\begin{figure}[htbp]
    \centering
    \begin{subfigure}{0.49\textwidth}
        \includegraphics[width=\textwidth]{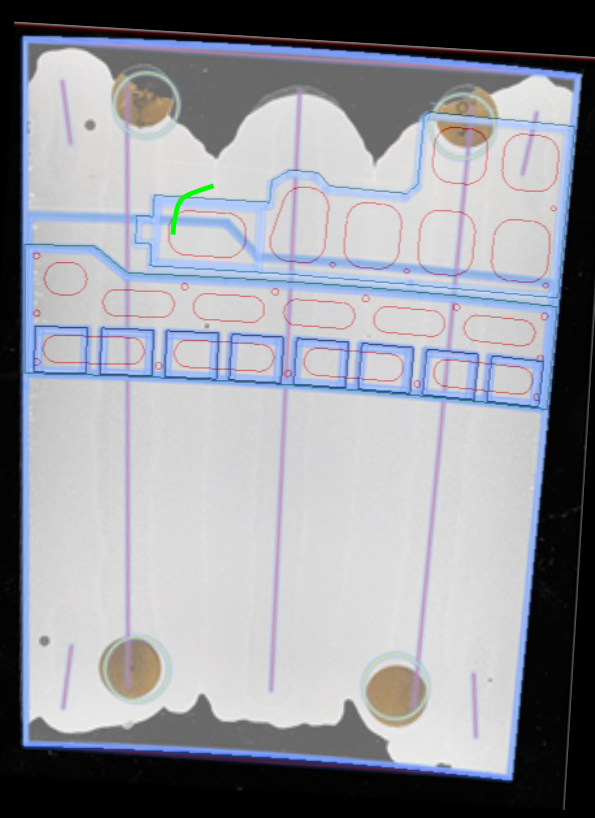}
        \caption{}
    \end{subfigure}
    \begin{subfigure}{\textwidth}
        \includegraphics[width=\textwidth]{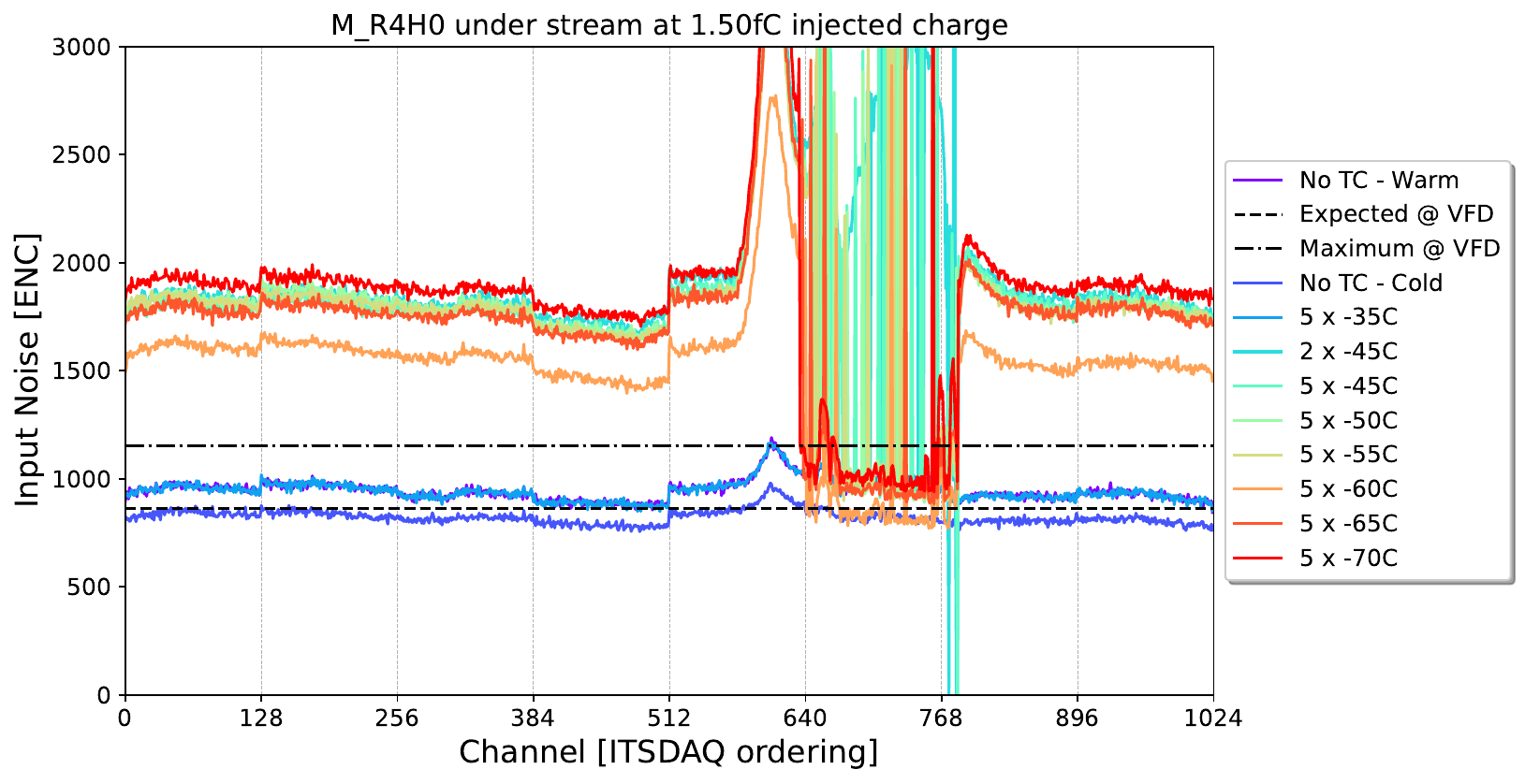}
        \caption{}
    \end{subfigure}
    \caption{(a) Full-coverage HYSOL pattern, shown in light grey, for an R4M0 module overlaid with the PCBs, shown in blue, and showing the PCB-to-sensor adhesive pattern in red stencil. The visually confirmed crack is shown in light green. (b) Noise results for the M\_R4M0 over all temperature cycles. Noise results after the $\unit[-55]{\oC}$ cycle are much higher on average due to a smaller bias voltage. Readout order is reversed with respect to physical sensor geometry.}
    \label{fig:hysol_fc_r4_crack}
\end{figure}

The S\_R2 and S\_R5 modules had sensor splinters missing at glue-dot locations, as expected. On the S\_R4 module, no crack was visually confirmed, while on the M\_R4 module, a crack was visually confirmed under the DC-DC converter that the PCB-to-sensor adhesive pattern. This crack was unexpected as the sensor was fully supported by HYSOL underneath and did not originate from a glue-dot which indicated that even a full-coverage pattern was insufficient to fully prevent cracking. Figure~\ref{fig:hysol_fc_r4_crack} shows the full-coverage loading pattern overlaid with the M\_R4M0 module and the location of the crack as well as the noise signature over multiple DAQ scans.

\subsubsection{Lessons learned from HYSOL optimized petal}
\label{subsubsec:hysol_fc_conclusion}

\begin{figure}[htbp]
    \centering
    \includegraphics[width=0.97\textwidth]{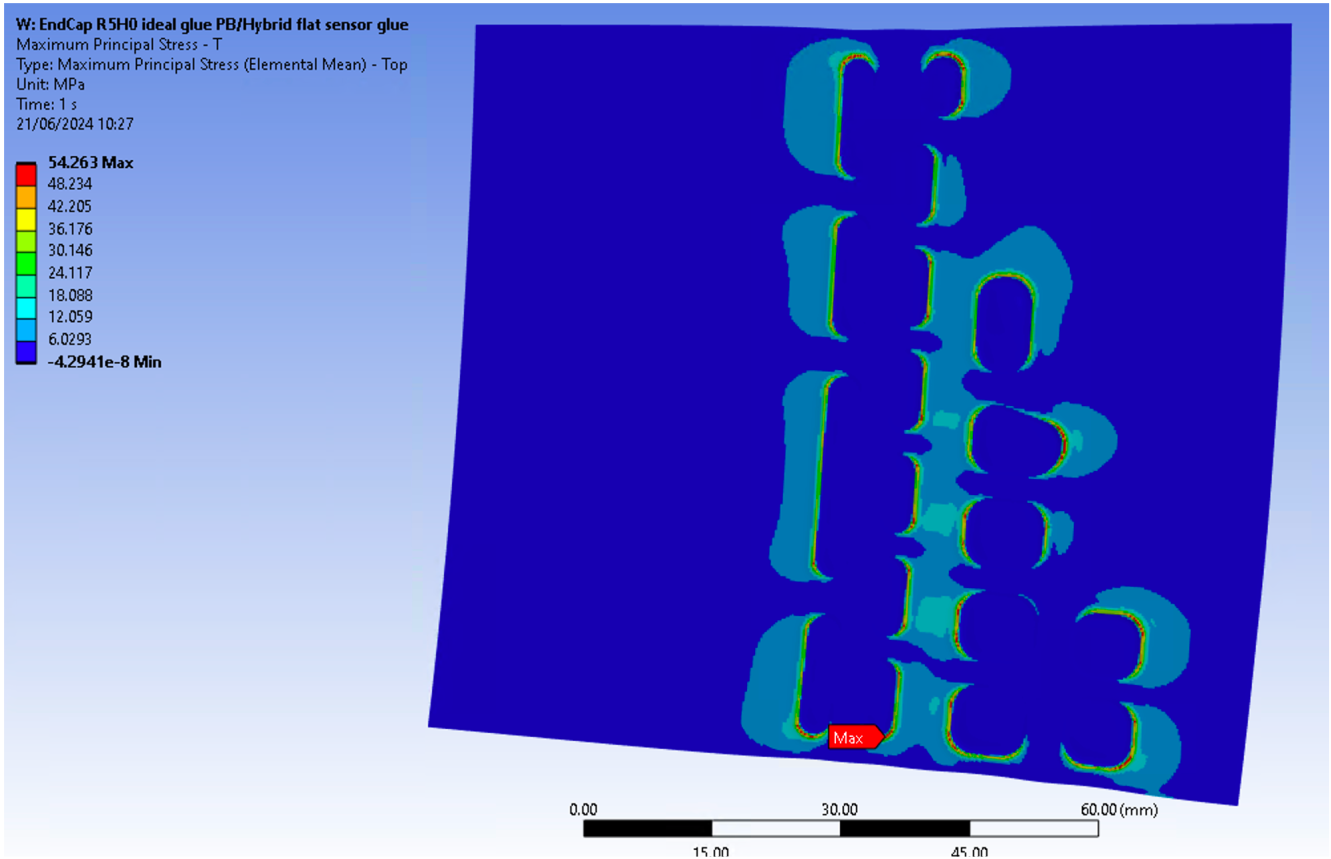}
    \caption{Stress simulation of an R5 module using HYSOL in a full-coverage loading pattern and a PCB-to-sensor pattern excluding glue-dots. The regions of highest stress occur at the edges of the PCBs, but the overall stress level of the module is reduced significantly.}
    \label{fig:hysol_fc_wo_dots_sim}
\end{figure}

Given that cracks originating from glue-dots were expected, another simulation was done using full-coverage HYSOL and a PCB-to-sensor adhesive pattern without glue-dots, as shown in figure~\ref{fig:hysol_fc_wo_dots_sim}. A significant reduction of stress is seen in comparison to figure~\ref{fig:hysol_fc_with_dots_sim}. This confirmed the locations of all the crack signatures except for the main side R4 module.

Moreover, given that HYSOL appears to be highly dependent on the loading pattern and resulted in an increased number of cracks compared to the default SE4445 petals (section~\ref{subsec:se4445_petals}), it was decided that HYSOL was not a reliable mitigation strategy for production. However, the studies with HYSOL in both the snake-like and full-coverage patterns led to several important conclusions:
\begin{itemize}
    \item The reliability of modules was improved significantly by complete coverage of PCB edges and gaps;
    \item The glue-dots used in the PCB-to-sensor adhesive pattern resulted in regions of high stress.
\end{itemize}
It was concluded that a combination of modified loading and module assembly adhesive patterns did indeed provide a significant stress reduction for modules.

\FloatBarrier

\subsection{HYSOL optimized and SE4445 default petal: DESY2}
\label{subsec:desy2}

The DESY2 petal began loading using HYSOL in a full-coverage pattern (section~\ref{subsec:hysol_fc}); however, due to the unexpected crack on the R4 module of the VAN4 petal, the ITk community decided to abandon the HYSOL studies and the loading adhesive and pattern were reverted to SE4445 in a snake-like pattern. The M\_R0, M\_R1, and M\_R3 modules were loaded with full-coverage HYSOL, as shown in figure~\ref{fig:hysol_fc_pattern}, while the remainder were loaded with SE4445 with a snake-like pattern but different from that used for the VAN2 (section~\ref{subsubsec:van2_result}) and IFIC2 (section~\ref{subsubsec:ific2_result}) petals. The DESY2 petal was first passively cycled to $\unit[-45]{\oC}$ and then actively cycled to $\unit[-75]{\oC}$. The initial cold test at $\unit[-35]{\oC}$ showed multiple early breakdowns for the M\_R1, M\_R3, M\_R5, and S\_R3M0 modules at bias voltages well below $\unit[-250]{V}$. Most DAQ scans were performed at a bias voltage of $\unit[-150]{V}$ or lower.

\begin{figure}[htbp]
    \begin{minipage}[b]{\linewidth}
    \centering
        \includegraphics[width=\textwidth]{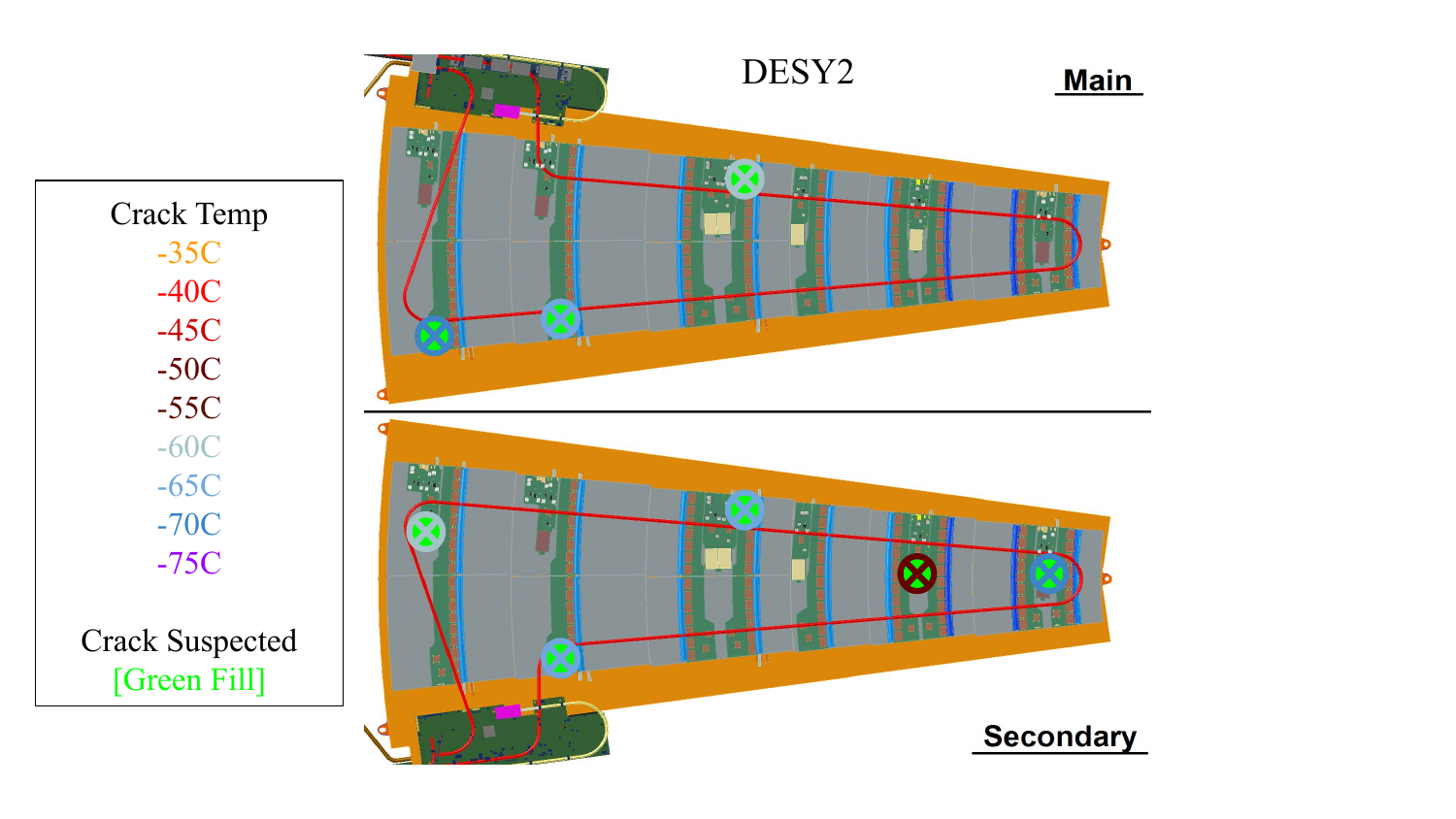}
    \end{minipage}
    \begin{minipage}[b]{\linewidth}
        \centering
        \begin{tabular}{ c | c | c || c | c | c}
            \toprule
            \shortstack{\textbf{Temperature} \\ \textbf{[$\oC$]}}         &
            \shortstack{\textbf{Module}      \\ \textbf{crack}}           &
            \shortstack{\textbf{Visually}    \\ \textbf{confirmed [Y/N]}} &
            \shortstack{\textbf{Temperature} \\ \textbf{[$\oC$]}}         &
            \shortstack{\textbf{Module}      \\ \textbf{crack}}           &
            \shortstack{\textbf{Visually}    \\ \textbf{confirmed [Y/N]}} \\
            \midrule
            $-50$ & S\_R1   & N &
            $-65$ & S\_R4M1 & N \\ \midrule
            $-60$ & M\_R3M0 & N &
            $-65$ & S\_R3M0 & N \\ \midrule
            $-60$ & S\_R5M0 & N &
            $-70$ & M\_R5M1 & N \\ \midrule
            $-65$ & M\_R4M1 & N &
            $-70$ & S\_R0   & N \\ \bottomrule
        \end{tabular}
    \end{minipage}
    \caption{Suspected crack locations on the DESY2 petal, after each crack signature was seen in the IV and DAQ scans.}
    \label{fig:desy2_overview}
\end{figure}

A total of eight crack signatures were seen, which are summarized in figure~\ref{fig:desy2_overview}. None of the eight crack signatures were visually confirmed, but three were near the DC-DC converters and the remaining at the edges, locations consistent with previously discovered cracks.

The S\_R1 crack, which occurred during thermal cycling at $\unit[-50]{\oC}$, was the first observation of a crack signature occurring in between the cold and warm cycle. The M\_R3M0 module showed similar behaviour to the S\_R1 module, with early breakdown occurring between the fourth cold cycle at $\unit[-60]{\oC}$ and the subsequent warm cycle. Similarly, the S\_R3M0 module exhibited a crack signature between the third and fourth cycles at $\unit[-65]{\oC}$.

\subsubsection{Lessons learned from DESY2 petal}
\label{subsubsec:DESY2_conclusion}

An interesting feature of the DESY2 petal was that the observed cracks were located partially in the same locations as the cracks on the HYSOL petal, even though fewer cracks were observed. Previously, cracks along the short PCB edges had only been observed if HYSOL was used for loading and had been attributed to the use of a harder adhesive for loading. Cracks on the DESY2 petal confirmed that this crack signature also occurred if SE4445 was used. Following this observation, the modulus of SE4445 at cold temperatures was tested in a differential material analysis measurement. The Young's modulus of SE4445 at temperatures below $\unit[-40]{\oC}$ was confirmed to be comparable to that of HYSOL, approximately \unit[2000]{MPa} at $\unit[-45]{\oC}$, explaining the similar crack signatures at cold temperatures.

Based on these observations, a new mitigation approach was designed, combining the major findings from previous petals:
\begin{itemize}
    \item Since small glue-dots in the PCB-to-sensor adhesive pattern were found to maximize stress, adhesive patterns for module assembly were designed to omit small glue-dots;
    \item HYSOL had not been found to reduce the number of cracks compared to petals loaded with SE4445, therefore the new loading approach used SE4445 again;
    \item Due to SE4445 developing a hardness comparable to HYSOL at cold temperatures, the pattern optimizations developed for loading with HYSOL --- covering gaps and edges --- were included in the new loading patterns along with curved adhesive lines to distribute the stress.
\end{itemize}
Since this approach contained optimizations of the default loading strategy, it is referred to as the ``Improved Nominal'' design.

\FloatBarrier

\subsection{SE4445 Improved Nominal petals: VAN6 and VAN7}
\label{subsec:impnom}

\begin{figure}[htbp]
    \centering
    \includegraphics[width=\textwidth]{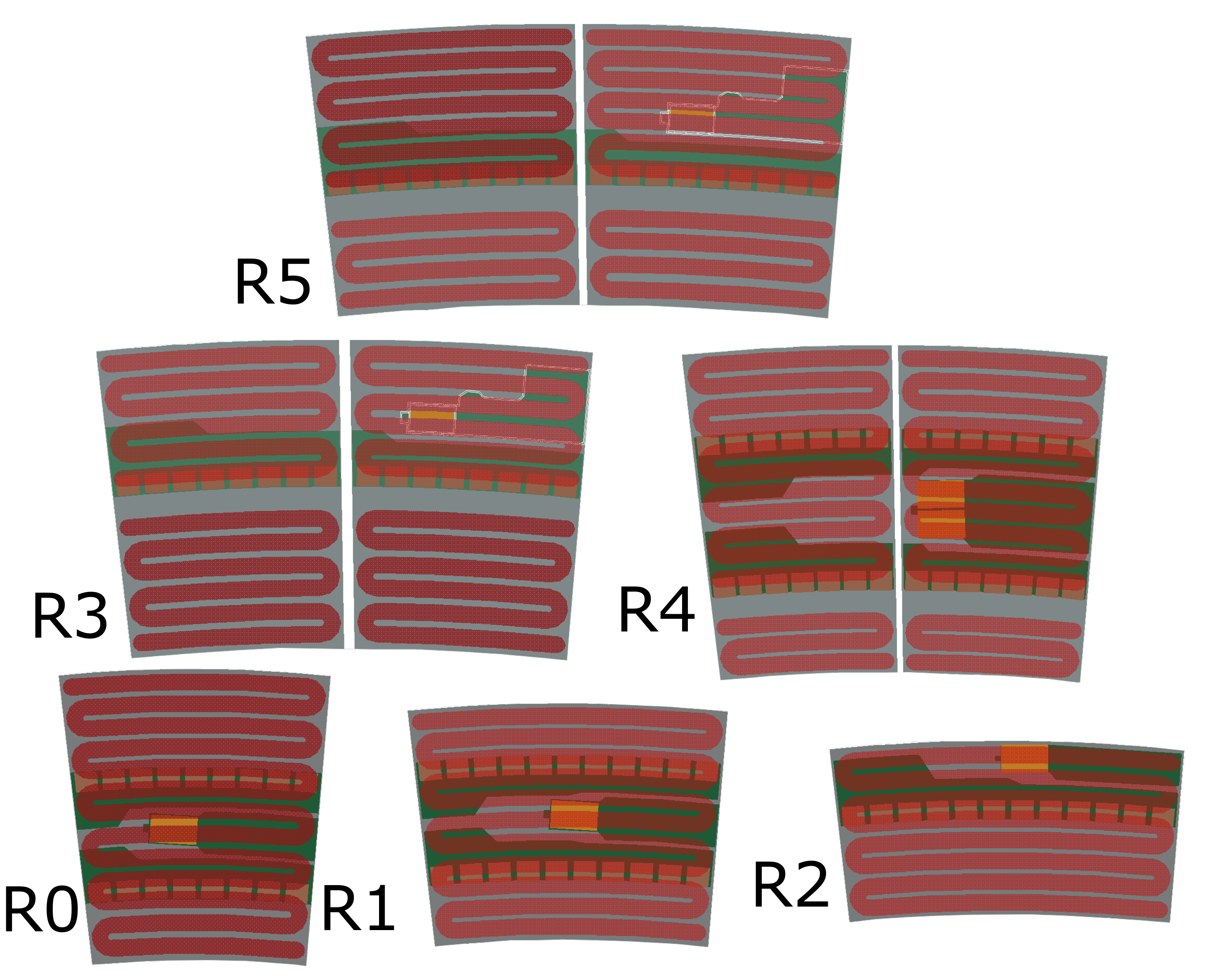}
    \caption{Loading adhesive patterns for the Improved Nominal design, in red, and overlaid with the PCBs, in green. The different module types are not necessarily to scale.}
    \label{fig:INglue}
\end{figure}

Two petals were constructed using the Improved Nominal assembly patterns for modules and module loading outlined in section~\ref{subsubsec:DESY2_conclusion}. Figure~\ref{fig:INglue} shows the improved loading pattern overlaid with the PCBs, highlighting the snake-like pattern and the high degree of coverage. In this paper, two petals loaded using the Improved Nominal design --- VAN6 and VAN7 --- are discussed.

\subsubsection{VAN6}
\label{subsubsec:van6_results}

The VAN6 petal was the first petal built using the Improved Nominal design. All modules were loaded within the specifications of placement and adhesive height with the exception of the S\_R1 and S\_R5 modules. The S\_R1 module was built with half the normal adhesive height, while the S\_R5 module was built with double the adhesive height. These out-of-specification modules were loaded due to a lack in part availability at the time but later served as a probe into the sensitivity of module adhesive height to cracking.

The VAN6 petal was passively cycled and tested to $\unit[-75]{\oC}$. The first crack signatures were observed on the out-of-specification modules after five cycles at $\unit[-65]{\oC}$. These crack signatures were consistent with the locations of the DC-DC converters observed on previous petals. As the crack signatures appeared at temperatures below the temperatures expected during the worst-case cooling failures, this demonstrates that there is space for deviations in module assembly, i.e.\ out-of-specification modules do not crack until $\unit[-65]{\oC}$, confirming that the Improved Nominal design is robust against variations in module assembly.

\begin{figure}[htbp]
    \begin{minipage}[b]{\linewidth}
        \centering
        \includegraphics[width=\textwidth]{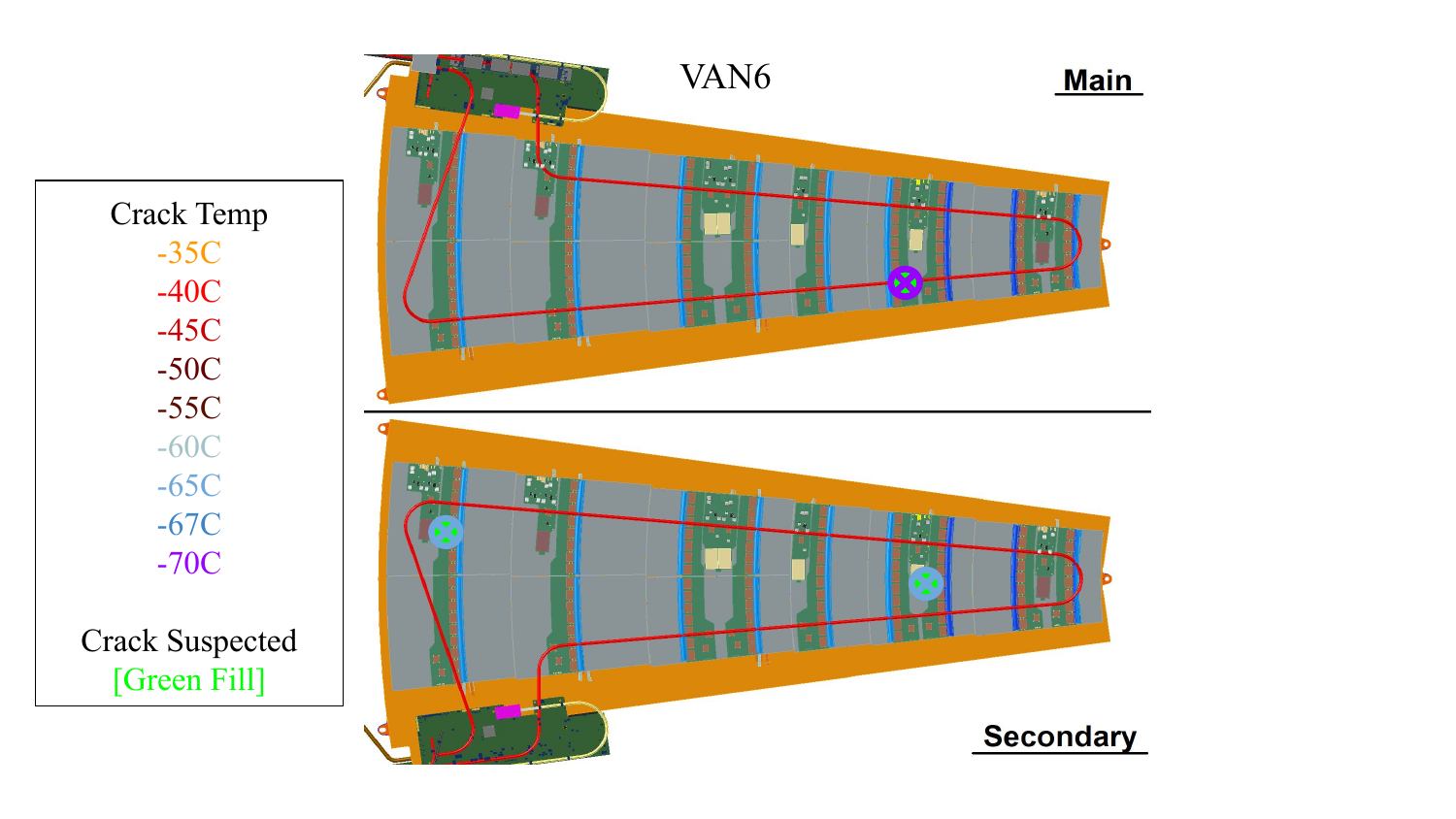}
    \end{minipage}
    \begin{minipage}[b]{\linewidth}
        \centering
        \begin{tabular}{ c | c | c || c | c | c}
            \toprule
            \shortstack{\textbf{Temperature} \\ \textbf{[$\oC$]}}         &
            \shortstack{\textbf{Module}      \\ \textbf{crack}}           &
            \shortstack{\textbf{Visually}    \\ \textbf{confirmed [Y/N]}} &
            \shortstack{\textbf{Temperature} \\ \textbf{[$\oC$]}}         &
            \shortstack{\textbf{Module}      \\ \textbf{crack}}           &
            \shortstack{\textbf{Visually}    \\ \textbf{confirmed [Y/N]}} \\
            \midrule
            $-65$ & S\_R1\dag   & N &
            $-75$ & M\_R1   & N \\ \midrule
            $-65$ & S\_R5M0\ddag & N &
                  &         &   \\ \bottomrule
        \end{tabular}
    \end{minipage}
    \caption{Suspected crack locations on the VAN6 petal, after each crack signature was seen in the IV and/or DAQ scans. (\dag{}) The S\_R1 module was built with half the normal adhesive height. (\ddag{}) The S\_R5 module was built with double the normal adhesive height.}
    \label{fig:van6_overview}
\end{figure}

The M\_R1 module had an early breakdown after cycling at $\unit[-75]{\oC}$; however, the only noise signature was a single channel with high noise. It is possible that this is a crack, but similar signatures have not been observed in previous cracked modules. Figure~\ref{fig:van6_overview} shows an overview of the suspected crack locations on the VAN6 petal.

\subsubsection{VAN7}
\label{subsubsec:van7_result}

\begin{figure}[htbp]
    \begin{minipage}[b]{\linewidth}
        \centering
        \includegraphics[width=\textwidth]{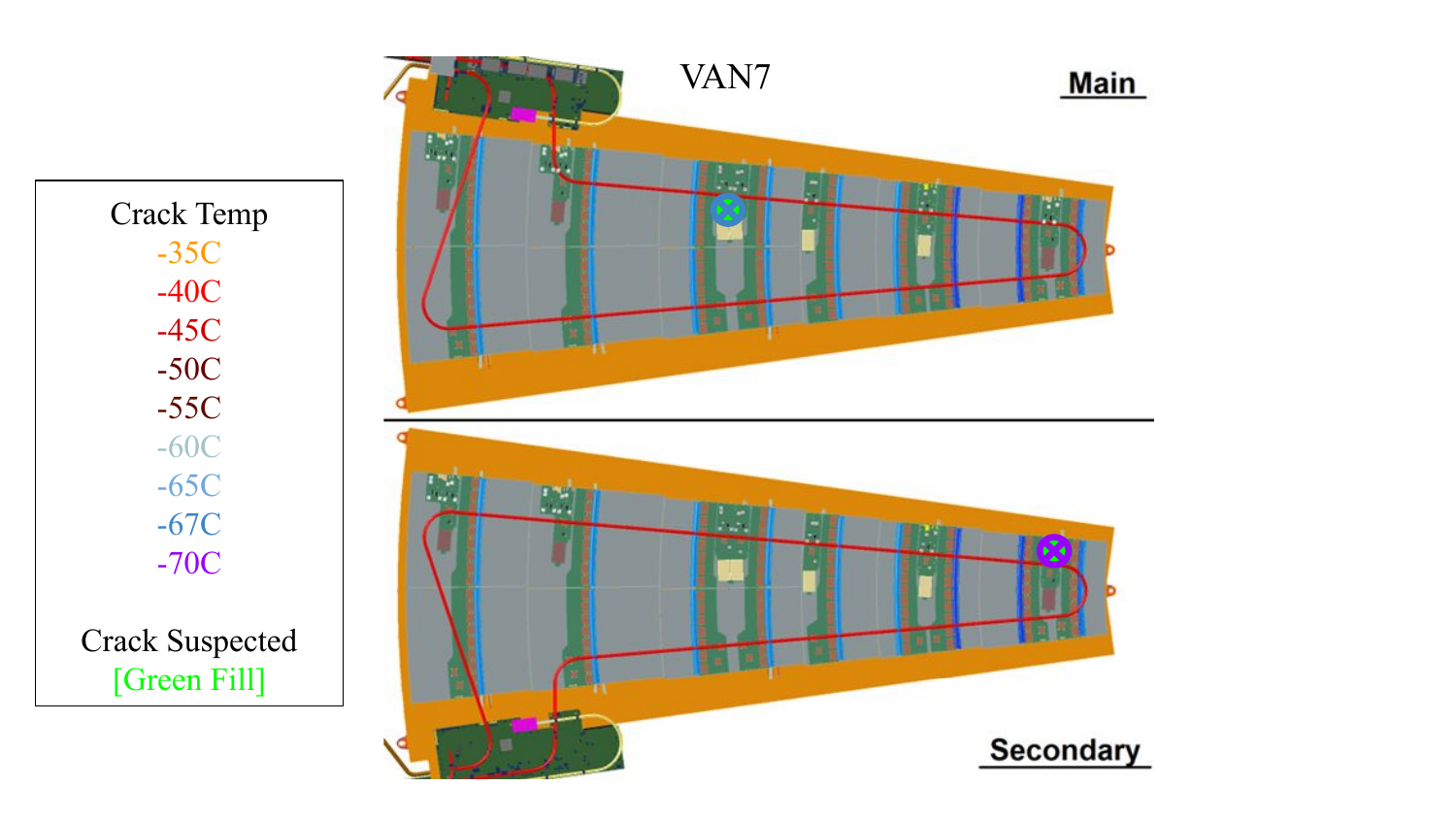}
    \end{minipage}
    \begin{minipage}[b]{\linewidth}
        \centering
        \begin{tabular}{ c | c | c || c | c | c}
            \toprule
            \shortstack{\textbf{Temperature} \\ \textbf{[$\oC$]}}         &
            \shortstack{\textbf{Module}      \\ \textbf{crack}}           &
            \shortstack{\textbf{Visually}    \\ \textbf{confirmed [Y/N]}} &
            \shortstack{\textbf{Temperature} \\ \textbf{[$\oC$]}}         &
            \shortstack{\textbf{Module}      \\ \textbf{crack}}           &
            \shortstack{\textbf{Visually}    \\ \textbf{confirmed [Y/N]}} \\
            \midrule
            $-67$ & M\_R3M1 & N &
            $-70$ & S\_R0   & N \\ \bottomrule
        \end{tabular}
    \end{minipage}
    \caption{Suspected crack locations on the VAN7 petal, after each crack signature was seen in the IV and/or DAQ scans.}
    \label{fig:van7_overview}
\end{figure}

The VAN7 petal was built with in-specification modules and using the Improved Nominal design. It was passively cycled and tested down to $\unit[-70]{\oC}$, testing only at warm temperatures. Due to a parts limitation, the M\_R3M0 module was loaded with a significant module bow and was used to investigate the effect of extremely bowed modules. An overview of the suspected crack locations is shown in figure~\ref{fig:van7_overview}. Neither of these two cracks were considered problematic due to their occurrence at temperatures unlikely to be reached in the detector; however, they are useful for determining the headroom of the Improved Nominal solution

\subsubsection{Further results of Improved Nominal petals}
\label{subsec:further_impnom}

After the Improved Nominal concept was developed and tested with the VAN6 and VAN7 petals, two additional petals were assembled following the same approach, the IFIC3 and IFIC4 petals. The IFIC3 petal was assembled from modules of the same build as those on the VAN6 and VAN7 petals and loaded using the Improved Nominal pattern. The IFIC3 petal cracked after thermal cycling at $\unit[-55]{\oC}$. The IFIC4 petal was assembled from modules with a modified adhesive pattern and loaded using the Improved Nominal pattern. The IFIC4 petal cracked after thermal cycling at $\unit[-45]{\oC}$. These additional petals underline the importance of the adhesive pattern for cracking mitigation.

Six additional petals were assembled using the Improved Nominal strategy; that is, the same module and loading adhesive patterns as the VAN6 and VAN7 petals. None of the modules exhibited early breakdown before or after loading. These six petals are part of the pre-series production of the ITk strip detector.

The testing for the six pre-series petals included an initial active cycle at +15 and $\unit[-35]{\oC}$ followed by passive cycling at $\unit[-45]{\oC}$ and subsequently at $\unit[-55]{\oC}$. Of the six, four petals cracked at a temperature between the $-45$ and $\unit[-55]{\oC}$ temperature cycles; the cracks are summarized in Table~\ref{tab:preseries_cracks}. The cracks are randomly distributed, with no preferred module type, and correspond to a cracking rate of a few percent. This highlights that under production conditions, the temperature headroom of the Improved Nominal strategy is reduced compared to the VAN6 and VAN7 petals but compatible with the detector requirement to have at most 10\% failing channels.

\begin{table}[htbp]
    \centering
    \caption{Suspected cracks on the pre-series petals built using the Improved Nominal mitigation strategy. All cracks occurred after thermal cycling at $\unit[-55]{\oC}$.}
    \label{tab:preseries_cracks}
    \begin{tabular}{c | c | l}
        \toprule
        Pre-series petal & Number of cracks & Module cracks \\
        \midrule
        1 & 1 & M\_R2 \\
        2 & 0 & -- \\
        3 & 2 & M\_R3M0 and M\_R3M1 \\
        4 & 1 & S\_R5M0 \\
        5 & 0 & -- \\
        6 & 1 & M\_R5M0 \\
        \bottomrule
    \end{tabular}
\end{table}

\FloatBarrier
\subsubsection{Lessons learned from Improved Nominal petals}
\label{subsubsec:se4445_inom_conclusion}

\begin{figure}[htbp]
    \centering
    \includegraphics[width=\textwidth]{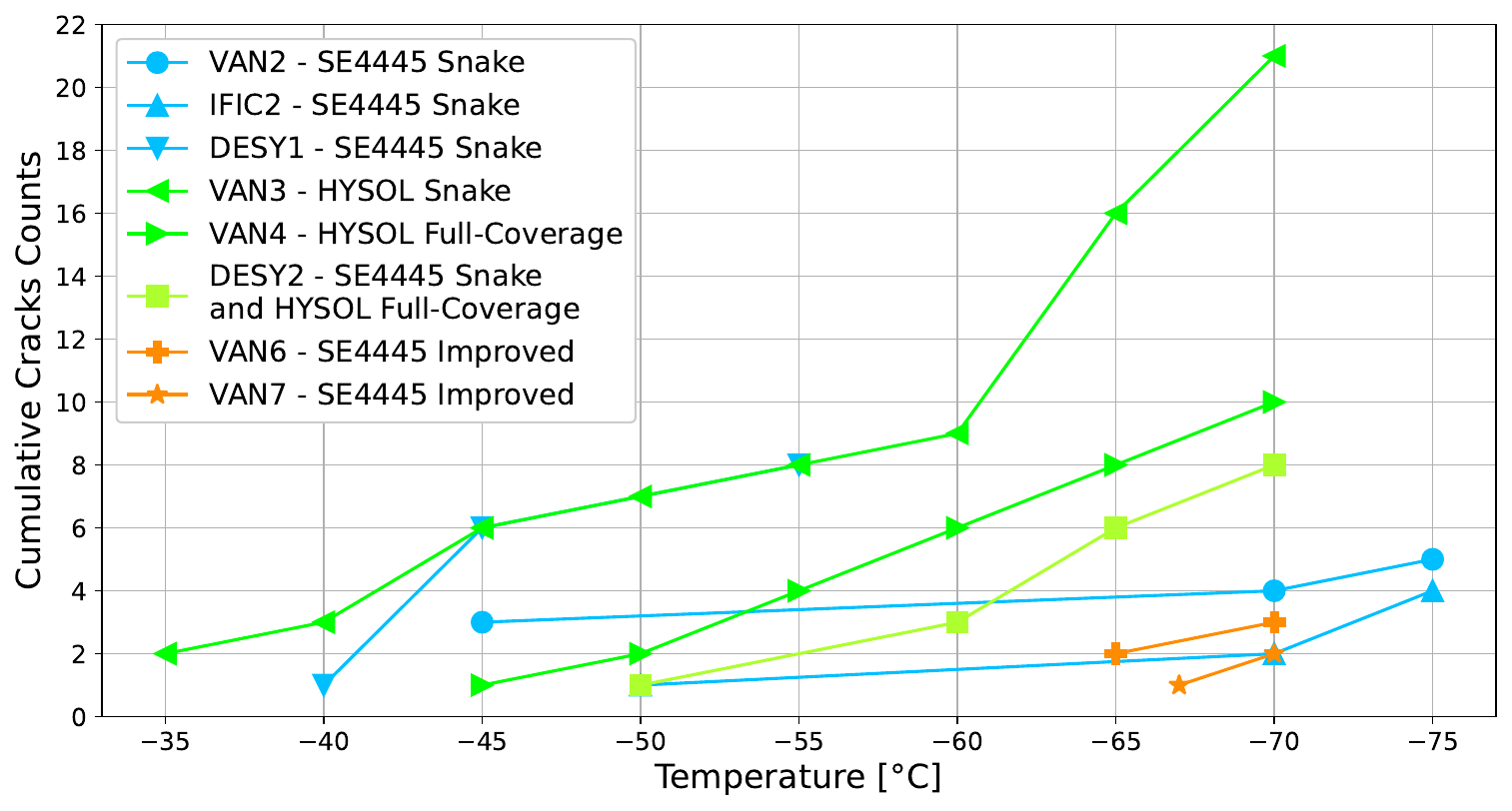}
    \caption{Summary of the number of (suspected plus observed) cracks across all mitigation petals. Comparing the onset of cracks in petals of the Improved Nominal design (orange) to that in petals without cracking mitigation (blue) or those utilizing HYSOL as a loading adhesive (green), there is approximately \unit[10]{\oC} of headroom provided by the Improved Nominal approach. Note that this plot does not include the additional results described in section~\ref{subsec:further_impnom}.}
    \label{fig:crack_summary}
\end{figure}

While cracks still formed on the Improved Nominal petals, the number of cracks and the temperature at which they first occurred was lower than for the previous petal designs. This confirmed that using SE4445 as the loading adhesive and removing the glue-dots from module adhesive pattern is the most promising mitigation strategy that does not change the module stackup to reduce stress. The temperatures at which cracks were observed in the Improved Nominal petals demonstrate that this mitigation provides increased headroom against the apparition of cracks compared to the default and the HYSOL mitigation strategies. Figure~\ref{fig:crack_summary} shows a cumulative summary of the cracks observed across mitigation petals.

\FloatBarrier

\FloatBarrier
\section{Conclusion}
\label{sec:conclusion}

To mitigate the cracking of ITk end-cap strip modules, an iterative process of changing adhesive types and patterns was investigated. The initial adhesive pattern with SE4445 exhibited cracks after thermal cycling at temperatures warmer than $\unit[-45]{\oC}$, temperatures which the detector may be exposed to during cooling failures.

HYSOL, a stiffer adhesive than SE4445, was tested as a loading adhesive to reduce the flexing of components and therefore stress. However, it was found that the use of HYSOL did not reduce the number of cracks or the temperature at which they first occur. This is because SE4445 has a similar stiffness to HYSOL at temperatures below $\unit[-35]{\oC}$ and because a non-ideal PCB-to-sensor adhesive pattern was used for module assembly on the petals using HYSOL. Therefore, this mitigation strategy was considered insufficient.

While several other SE4445 loading patterns were tested, it was the removal of the glue-dots in the module adhesive pattern that most drastically reduced the rate of cracking, especially at temperatures closest to the nominal operational temperature of $\unit[-35]{\oC}$. This was demonstrated in the studies of the Improved Nominal design, where cracking only occurred after thermal cycling at temperatures colder than $\unit[-45]{\oC}$. Cracks at these temperatures allow for additional headroom from the normal operating temperature of the ITk. The Improved Nominal design was therefore considered a viable temporary approach for the initial phase of petal construction and is being used for the construction of part of the end-cap detectors, complementing the interposer approach adopted for most petals.

\FloatBarrier
\acknowledgments
\addcontentsline{toc}{section}{Acknowledgements}

This work has been supported by the Natural Sciences and Engineering Research Council of Canada (NSERC), the Canada Foundation for Innovation (CFI), the German Federal Ministry of Education and Research (BMBF), the Helmholtz Association of German Research Centres (HGF), the Spanish Ministry of Science, Innovation and Universities (MICINN), and the Spanish State Research Agency (AEI).

\bibliographystyle{JHEP}
\bibliography{bibliography}

\end{document}